\documentclass[fleqn,usenatbib]{mnras}

\usepackage{newtxtext,newtxmath}

\usepackage[T1]{fontenc}
\usepackage{ae,aecompl}

\usepackage{graphicx}	
\usepackage{amsmath}	
\usepackage{amssymb}	
\usepackage{color}
\usepackage{gensymb}

\newcommand\uJy{$\mu$Jy}
\newcommand\eMERLIN{$e$-MERLIN}

\title[SuperCLASS -- II: Photometric Redshifts]{SuperCLASS -- II: Photometric Redshifts and Characteristics of Spatially-Resolved \uJy\ Radio Sources}

\author[S. M. Manning et al.]{
Sinclaire M. Manning$^{1}$,\thanks{E-mail: smanning@astro.as.utexas.edu}
Caitlin M. Casey$^{1}$,
Chao-Ling Hung$^{2}$,
Richard Battye$^{3}$,
\newauthor Michael L. Brown$^{3}$,
Neal Jackson$^{3}$,
Filipe Abdalla$^{4}$,
Scott Chapman$^{5}$,
\newauthor Constantinos Demetroullas$^{3,6}$,
Patrick Drew$^{1}$,
Christopher A. Hales$^{7,8}$,
\newauthor Ian Harrison$^{3,9}$,
Christopher J. Riseley$^{10,11,12}$,
David B. Sanders$^{13}$, 
\newauthor and Robert A. Watson$^{3}$
\\
$^{1}$Department of Astronomy, The University of Texas at Austin, 2515 Speedway Boulevard Stop C1400, Austin, TX 78712, USA \\
$^{2}$Physics Department, Manhattan College, 4513 Manhattan College Pkwy, Bronx, NY 10471, USA \\
$^{3}$Jodrell Bank Centre for Astrophysics, Department of Physics and Astronomy, The University of Manchester, Manchester M13 9PL, UK \\
$^{4}$Department of Physics and Astronomy, University College London, Gower Place, London WC1E 6BT, UK\\
$^{5}$Department of Physics and Atmospheric Science, Dalhousie University, Halifax, Nova Scotia, Canada\\
$^{6}$Cyprus University of Technology, Archiepiskopou Kyprianou 30, Limassol, 3036, Cyprus\\
$^{7}$National, Radio Astronomy Observatory, P.O. Box 0, Socorro, NM 87801, USA\\
$^{8}$School of Mathematics, Statistics and Physics, Newcastle University, Newcastle upon Tyne NE1 7RU, UK\\
$^{9}$Department of Physics, University of Oxford, Denys Wilkinson Building, Keble Road, Oxford OX1 3RH, UK\\
$^{10}$Dipartimento di Fisica e Astronomia, Universit\`a degli Studi di Bologna, via P. Gobetti 93/2, 40129 Bologna, Italy\\
$^{11}$INAF -- Istituto di Radioastronomia, via P. Gobetti 101, 40129 Bologna, Italy\\
$^{12}$CSIRO Astronomy and Space Science, PO Box 1130, Bentley, WA 6102, Australia\\
$^{13}$Institute for Astronomy, 2680 Woodlawn Drive, University of Hawaii, Honolulu, HI 96822, USA 
}

\date{Accepted XXX. Received YYY; in original form ZZZ}

\pubyear{2019}

\begin{document}
\label{firstpage}
\pagerange{\pageref{firstpage}--\pageref{lastpage}}
\maketitle

\begin{abstract}
We present optical and near-infrared imaging covering a $\sim$1.53\,deg$^2$ region in the Super-Cluster Assisted Shear Survey (SuperCLASS) field, which aims to make the first robust weak lensing measurement at radio wavelengths. We derive photometric redshifts for $\approx$176,000 sources down to $i^\prime_{\rm AB}\sim24$ and present photometric redshifts for 1.4\,GHz \eMERLIN\ and 
VLA detected radio sources found in the central 0.26\,deg$^{2}$. 
We compile an initial catalog of 149 radio sources brighter than S$_{1.4}>75$\,\uJy\ and find their photometric redshifts span $0<z_{\rm phot}<4$ with radio luminosities between $10^{21}-10^{25}$\,W\,Hz$^{-1}$, with medians of $\langle z \rangle \,=0.55$ and $\langle L_{1.4}\rangle \,=1.9\times10^{23}$\,W\,Hz$^{-1}$ respectively. We find 95\% of the \uJy\ radio source sample (141/149) have SEDs best fit by star-forming templates while 5\% (8/149) are better fit by AGN. Spectral indices are calculated for sources with radio observations from VLA and GMRT at 325\,MHz, with an average spectral slope of $\alpha=0.59\pm0.04$. Using the full photometric redshift catalog we construct a density map at the redshift of the known galaxy clusters, $z=0.20\pm0.08$. Four of the five clusters are prominently detected at $>7\,\sigma$ in the density map and we confirm the photometric redshifts are consistent with previously measured spectra from a few galaxies at the cluster centers.
\end{abstract}




\section{Introduction}
\label{sec:intro}
As direct probes of obscured star formation, infrared (IR) through radio wavelength observations provide crucial insight into the formation and evolution of galaxies across cosmic time. Galaxies with IR luminosities around $10^{12}\sim10^{13}$ L$_{\odot}$ from 8--1000$\mu$m have estimated star-formation rates (SFRs) of hundreds to thousands of solar masses per year and represent the most intense starbursts in the Universe \citep[e.g.][]{Smail1997,Chapman2005}. Surveys from the {\it Herschel Space Observatory} tell us that these galaxies contribute significantly to the total cosmic star-formation history, providing at least half of the Universe's star-formation at its peak at $z\sim2$ \citep{Casey2012,Gruppioni2013,Madau2014}.

The population of dusty star-forming galaxies \citep*[DSFGs;][]{Casey2014} relates closely to the \uJy\ radio source population \citep*[e.g.][]{Chapman2003,Barger2007}. For the purposes of this work \uJy\ radio sources are galaxies with 1.4\,GHz flux densities between 10--1000\,\uJy\ whose radio emission is most likely dominated by synchrotron emission from supernova remnants, which closely tracks their star-formation rates \citep{Helou1985}. However, placing meaningful constraints on the physical origins and drivers of the intense star formation in these systems has been difficult. This is due to the significant obscuration at optical/near-infrared (OIR) wavelengths caused by dust formed in these galaxies' star-forming regions \citep{Magnelli2009,Whitaker2017}. For example, it is still unclear from both observations \citep{Tacconi2008,Engel2010,Ivison2012,Hodge2012} and theory \citep{Dekel2009,Dave2010,Narayanan2010,Narayanan2015} whether this enhanced star formation in DSFGs/\uJy\ radio sources is triggered by a steady build up of material via cold gas accretion or short bursts via mergers.


The expanded Multi-Element Radio Linked Interferometer Network (\eMERLIN) is a network of seven UK-based radio telescopes designed for high spatial resolution $\sim$\,GHz observations. With a maximum baseline of 217\,km, corresponding to a resolution of $\approx200$\,mas at 1.4\,GHz, the facility's unique combination of sensitivity and spatial resolution makes it an ideal tool to trace spatially resolved star-formation in heavily obscured galaxies like the \uJy\ radio source population. \eMERLIN\ is the only radio telescope capable of resolving the internal, unobscured star formation within a statistically significant population of high-redshift galaxies as it is a dedicated long-baseline facility. \eMERLIN\ has embarked on an ambitious $\sim$\,1\,deg$^{2}$, high-resolution (0.3\arcsec), 1.4\,GHz (L-band) radio continuum survey of a new extragalactic deep field, called the Super-Cluster Assisted Shear Survey (SuperCLASS) using 832 hours of observations.

SuperCLASS (Paper I in this series) is a deep, wide-area survey using the \eMERLIN\ interferometric array designed to detect the effects of weak lensing in radio continuum in a super-cluster region where there are five $z\sim0.2$ Abell galaxy clusters already identified \citep{Abell1989,Struble1999}. The SuperCLASS survey's primary science goals are to: (1) provide a test-bed for weak lensing studies at radio wavelengths for the future Square Kilometer Array (SKA) and other SKA progenitors and (2) obtain internal $\sim$\,kpc maps of \uJy\ radio sources for statistically large samples, determining their evolutionary origins through morphological analysis. High resolution \eMERLIN\ imaging combined with observations from the Karl G. Janksy Very Large Array (VLA) provides results similar to that expected of the SKA, allowing for the development of the tools required for shape measurement and a quantitative assessment of the physical properties of radio sources that can be used for cosmic shear measurements \citep{Brown2011,Harrison2016,Bonaldi2016,Camera2017}.

This paper presents the initial catalog of photometric redshifts as measured via new optical and near-IR imaging data for SuperCLASS obtained from Subaru and {\it Spitzer}. With these data, we present an initial analysis of radio-detected galaxies in the field and the distribution of galaxies in and around the five $z\sim0.2$ Abell galaxy clusters. We outline the individual data sets compiled for the survey in Section \ref{sec:data_and_obs} and we describe the methods used to measure photometric redshifts in Section \ref{sec:phot-z}. We discuss the redshift distribution of radio sources and distribution of sources surrounding the field's galaxy clusters are presented in Section \ref{sec:results}. Section \ref{sec:summary} summarizes our results. We assume a $Planck$ $\Lambda$CDM cosmology with $\Omega_m=0.307$ and $H_0=67.7$\, km\,s$^{-1}$\,Mpc$^{-1}$ \citep{Collaboration2016a}.

\section{Data and Observations}
\label{sec:data_and_obs}
\begin{figure}
\includegraphics[width=\columnwidth]{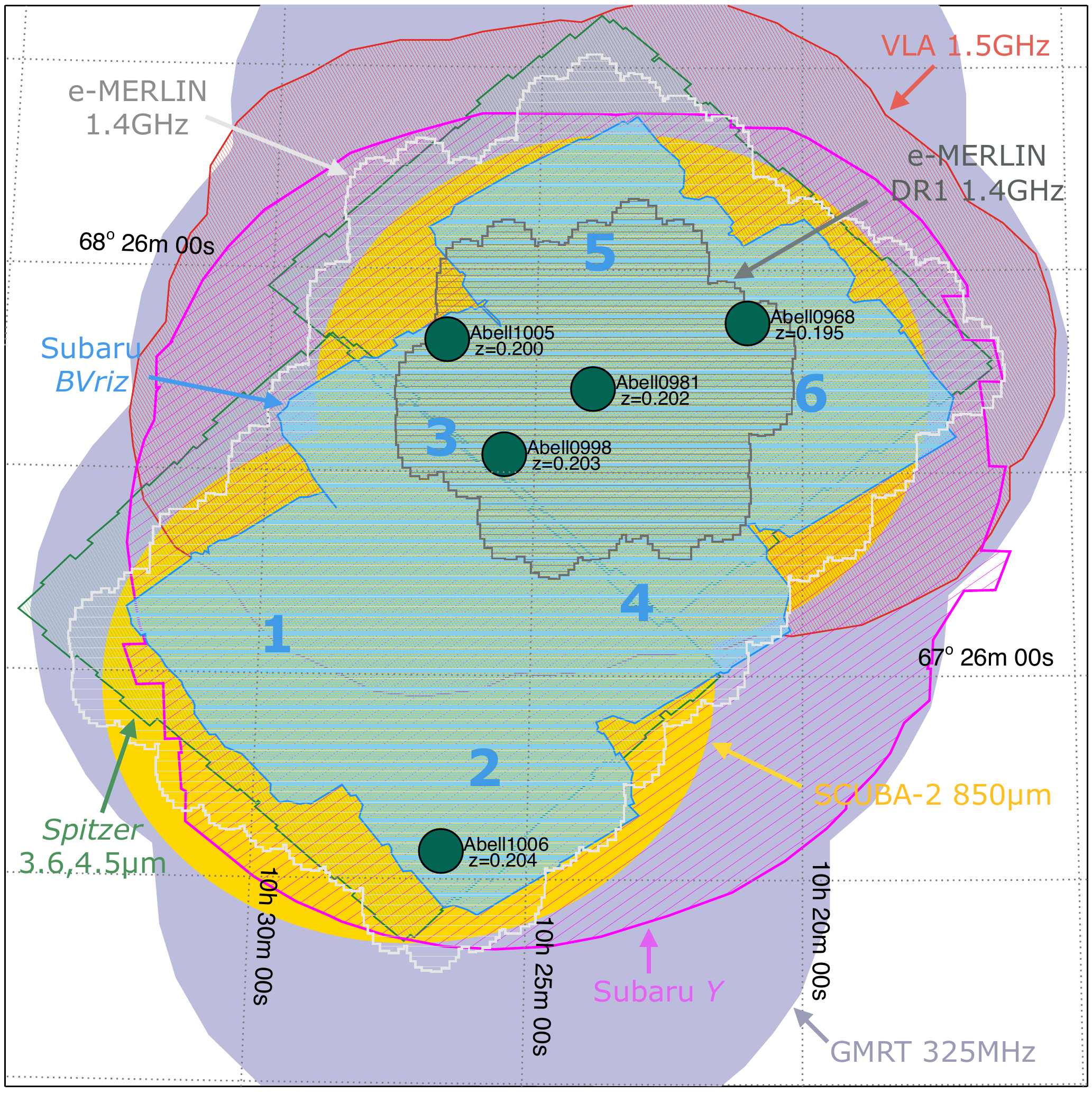}
\caption{SuperCLASS coverage map showing all collected observations, plus the full proposed \eMERLIN\ field (white). The six footprints covered by Subaru Suprime-Cam \textit{BVriz} are shown in blue, Subaru HyperSuprime-Cam \textit{Y}-band in magenta, \textit{Spitzer} $3.6,4.5\,\micron$ data in green, SCUBA-2 850\,\micron\ in yellow, GMRT 325\,MHz data in lavender, the \eMERLIN\ 1.4\,GHz DR1 in dark gray, and VLA 1.5\,GHz in red.} 
\label{fig:SC_map}
\end{figure}

What follows is a description of the core data products of the SuperCLASS survey, from the optical through near-IR, to radio observations. The optical and near-IR data are used to estimate photometric redshifts while the \eMERLIN/VLA data is used for a discussion of physical characteristics of the radio luminous sample. For a more thorough overview of the radio survey datasets in SuperCLASS, we refer the reader to Paper I (\eMERLIN, VLA) and \citet{Riseley2016} for the GMRT data. Paper III provides an initial analysis of the weak lensing signal in the field. Figure \ref{fig:SC_map} provides a visual summary of the SuperCLASS survey as it currently stands as well as the positions of the five Abell clusters, which together constitute a supercluster at $z\sim0.2$.

\subsection{Subaru Optical Imaging}
\label{subsec:subaru} 
Deep photometric data in the optical bands \textit{BVriz} were obtained using Suprime-Cam (SC) on the Subaru 8.4\,m telescope on Maunakea on 2013 February 5 and 2013 February 6 (PI Casey). Most of the second night (Feb 6) was lost due to poor weather and cloud cover. Seeing ranged from 1.0--1.4\arcsec. The SC data is comprised of six observed sub-fields, each with a FOV of $34\arcmin\times27\arcmin$, covering 1.53\,deg$^2$ total. Individual and total exposure times in these bands were: 200 seconds, totaling 1000 seconds for all six footprints ($B$ and $V$); 150 seconds, totaling 750 seconds for all six footprints ($r^\prime$); 200 seconds, totaling 2000 seconds for all six footprints ($i^\prime$); 240 seconds, totaling 1200 seconds for footprints 1-3 and 361.5 seconds for footprint 4 ($z^\prime$). Table \ref{tab:subaru_footprints} gives the position centroids for the six footprints and their position angles east of north about the centroid.

\begin{table}
\caption{Suprime-Cam Footprint Info}
\label{tab:subaru_footprints}
\begin{tabular}{cccc}
\hline
Footprint & RA & DEC & P.A. \\
\hline
1 & 10:30:38.8 & +67:30:18.2 & 33.5$\degree$ \\
2 & 10:27:07.1 & +67:16:00.2 & 33.8$\degree$ \\
3 & 10:27:23.5 & +67:56:35.9 & 33.3$\degree$ \\
4 & 10:23:50.9 & +67:42:14.4 & 33.3$\degree$ \\
5 & 10:24:04.7 & +68:22:46.6 & 33.9$\degree$ \\
6 & 10:20:31.2 & +68:08:21.5 & 33.7$\degree$ \\
\hline
\end{tabular}

{\bf Note --} Position centroids and position angles of the six Subaru Suprime-Cam footprints.
\end{table}

\textit{Y}-band imaging was obtained using Hyper Suprime-Cam (HSC) on 2015-March-27 (PI Hung). HSC has a 90\arcmin\ diameter FOV. Total integration time for the northeast image was 106 minutes and 96 minutes for the southwest. Seeing varied between 0.61-1.25\arcsec\ for the individual HSC exposures. There are $\sim$195,000 sources with $Y<24.3$ which we deem as a reliable counterpart match to our catalog generated from SC data, to be discussed later in this section. The inclusion of this filter proves useful as it probes key colors of both the $z=0.2$ population as well as the high-$z$ tail in which we are interested.

Due to the differences in seeing, conditions, and exposure times between the \textit{BVrizY} optical images, we convolve all images with a matched point-spread function (PSF) to perform multi-band photometry and eliminate all resolution discrepancies. PSF matching is crucial for accurate estimations of colors \citep{Capak2007} as otherwise aperture differences can significantly impact photometric redshift estimation. To do this, we use the SC image with the worst seeing (1.38\arcsec, full-width half maximum (FWHM)) as our reference PSF and create a PSF kernel. The kernel is then convolved with the remaining SC and HSC images to match the PSF of the reference image following standard procedure.

We use Source Extractor \citep[{\tt SExtractor};][]{Bertin1996} to build a catalog of objects from the PSF matched Subaru images and extract positions, magnitudes, and associated errors of the sources in the field. {\tt SExtractor} subtracts the estimated background from the photometry and uses the root mean square (RMS) to estimate errors. We find the uncertainties to be underestimated by a factor of ten based on the noise in the images and thus add a floor of 0.01 to the errors to account for this. We apply an aperture size with a 15 pixel diameter (3\arcsec), a minimum detection area of 3 pixels, and 3\,$\sigma$ detection threshold in the {\tt SExtractor} parameters to detect sources. We run {\tt SExtractor} in dual-image mode in which a deep, co-added image from all optical filters is used for object detection \citep{Capak2007,Laigle2016}. Forced photometry is then measured from each individual filter image. This procedure allows for the same aperture to be applied across all filters and objects are automatically matched, providing straightforward comparisons and cross-matching between different bands. A background RMS map with a mesh size of 12.8\arcsec\ is created and applied allowing for {\tt SExtractor} to reduce the number of spurious detections.

The northernmost 1/3 of the field lacks $z^\prime$-band data due to on the fly decisions made while observing; at the time, the \eMERLIN\ coverage area was planned to extend across the full 1.53\,deg$^2$. 
Unfortunately, this 1/3 of the field encompasses where the deepest \eMERLIN\ map sits. The impact of lack of $z^\prime$-band coverage on the reliability of the photometric redshifts is discussed further in Section \ref{sec:phot-z}.

The Subaru images are then flux calibrated using known stars in the large field of view. The FOV of both SC and HSC datasets is sufficiently large to have standard stars within our science frames. 
We make use of the Data Release 1 (DR1) Pan-STARRS (Panoramic Survey Telescope And Rapid Response System) 3$\pi$ survey \citep{Chambers2016} as our astrometry and photometry reference down to AB\,$<$\,22. Non-linear astrometric warps are corrected with astrometry.net \citep{Lang2009}. To flux calibrate we take a sample of 1,192 unsaturated stars across all six footprints in both SC/HSC and PS1 3$\pi$ with magnitudes between 15-19. We color correct from PS1 \textit{grizy} to our \textit{BVrizY} bands using stellar templates from the Stellar Spectral Flux Library \citep{Pickles1998}, convolving them with the spectral response functions (including atmosphere, detector throughput, and quantum efficiency) of the Subaru filters, and compare their magnitudes across filter sets. We then identify the best fit stellar models for each observed star using a $\chi^2$ minimization technique and determine a global offset to be applied to flux calibrate. This is folded into our catalog generation via the \texttt{MAG\_ZEROPOINT} parameter in the {\tt SExtractor} configuration file. 

\begin{figure*}
\begin{minipage}[t]{0.45\textwidth}
\includegraphics[width=\textwidth]{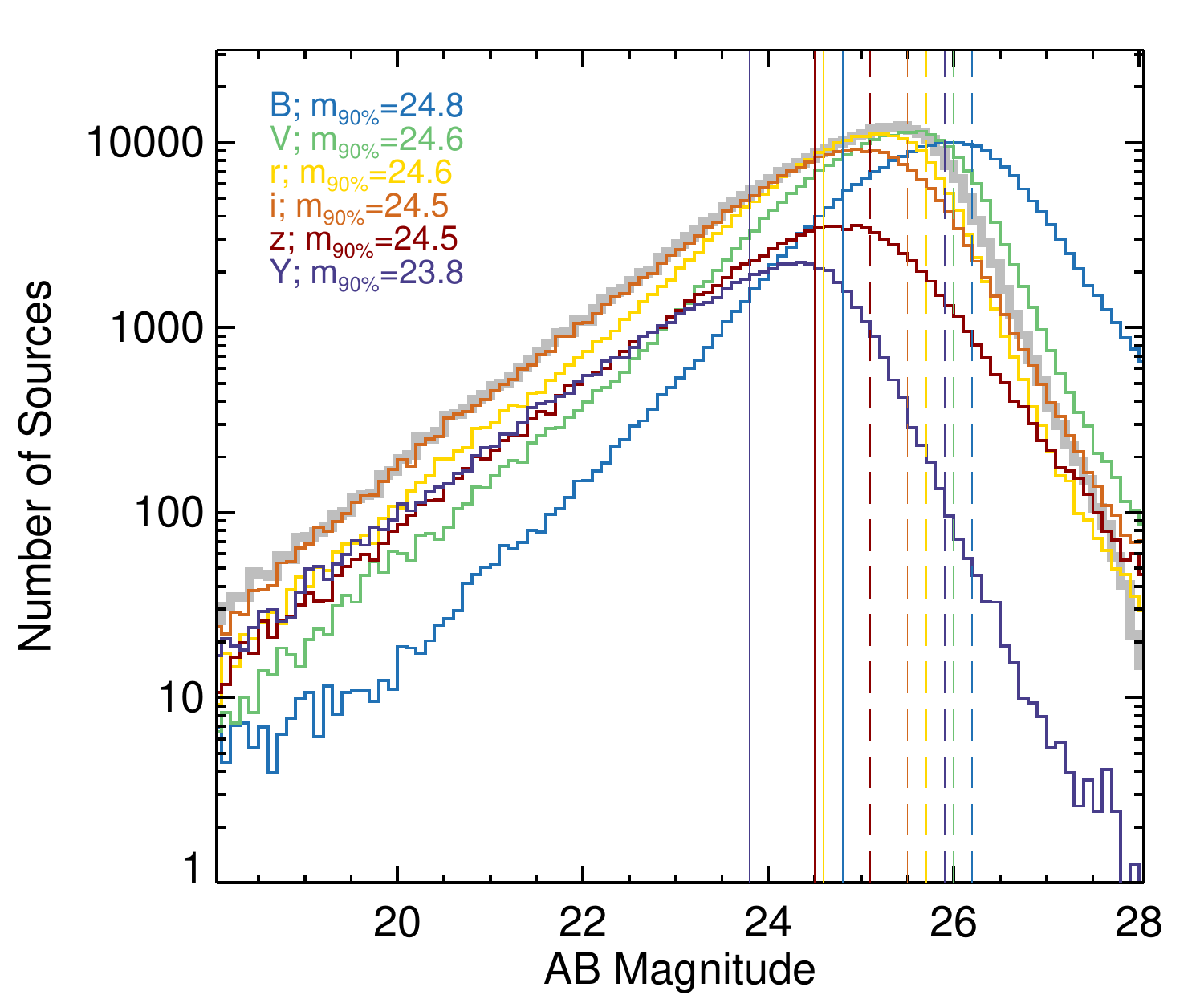}
\end{minipage}
\begin{minipage}[t]{0.45\textwidth}
\includegraphics[width=\textwidth]{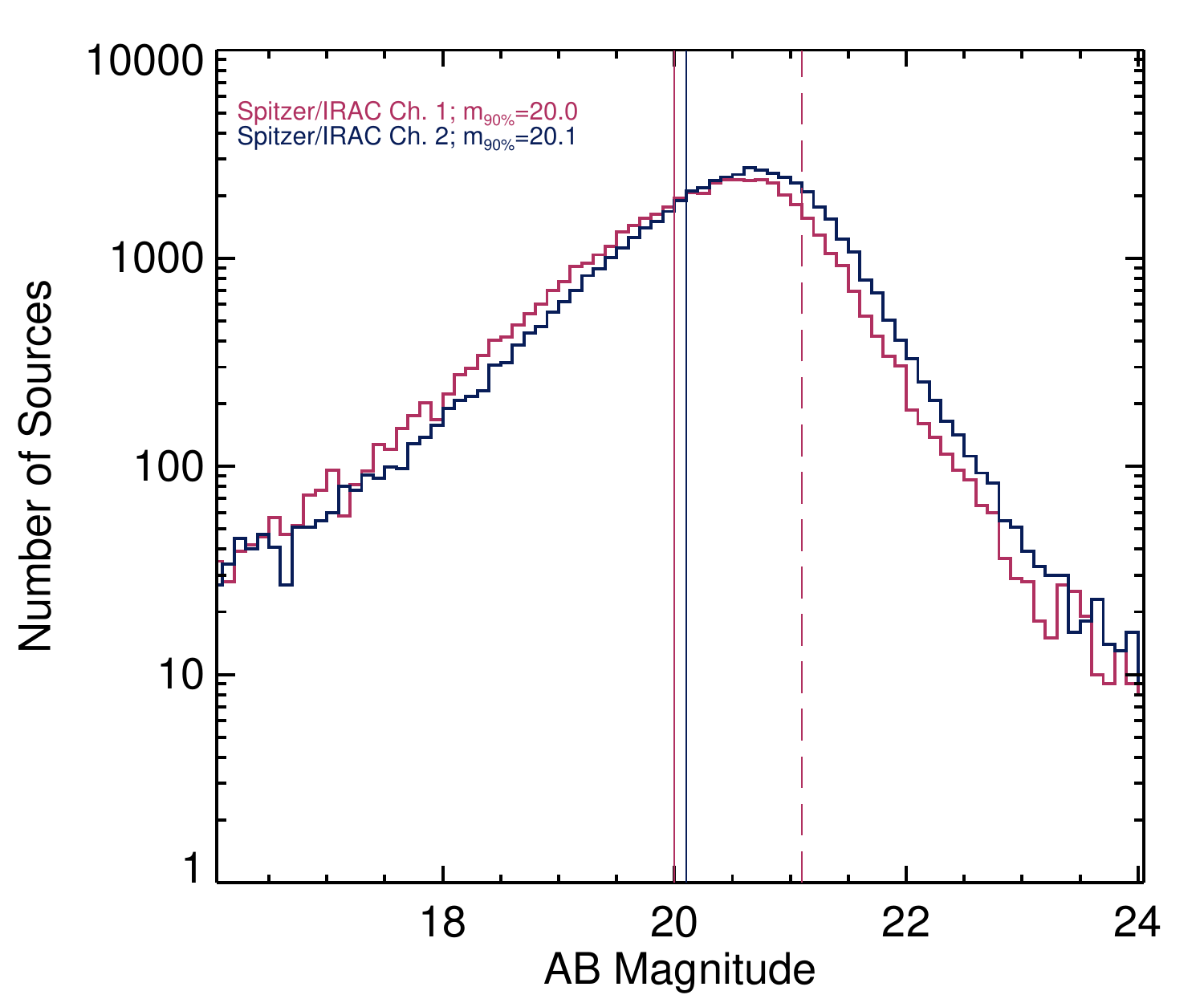}
\end{minipage}
\caption{Aggregate differential number counts showing image depth and completeness for the optical Subaru \textit{BVrizY} (left) and \textit{Spitzer} $3.6\,\micron$ and $4.5\,\micron$ observations (right). Vertical solid lines show at what depth we are 90\% complete in our sample in each photometric filter and dashed lines show the $5\,\sigma$ detection thresholds. The thick gray line shows the $i^\prime$-band number count using the single filter image rather than the co-added detection image. The \textit{Spitzer} 3.6 and 4.5\,$\micron$ 5\,$\sigma$ detection limits are both 21.1 mag.}
\label{fig:completeness_plots}
\end{figure*}

Some sources may be detected in more than one Subaru image due to spatial overlap between the six footprints. We identify duplicate detections and calculate the inverse-variance weighted average of the magnitudes so as to make use of all available data. This results in a master catalog of 376,380 objects, but after downselecting based on the $i^{\prime}$-band 90\% completeness limit ($i^{\prime}<24.5$), we end up with 176,447 sources. This work broadly agrees with expected optical number counts \citep{Nagashima2002}. Figure \ref{fig:completeness_plots} shows the 90\% completeness limits and 5\,$\sigma$ detection thresholds for these optical and near-IR observations. We calculate our completeness limits using a linear approximation of log(N) extrapolating down to 90\%. The use of a co-added detection image during source extraction results in an increased number of faint object detections and therefore we do not see the steep drop off in the number counts that is expected. The thick gray line shows the $i^\prime$-band number count without using a detection image and here we recover the typical profile. We note that the single-filter $i^\prime$-band images are noisier compared to the smooth, psf-matched co-added image. As such, we boost the detection threshold and minimum detections area to $5\,\sigma$ and 10\,pixels respectively to decrease the number of contaminants. There still exists an excess in the single filter number counts between 25-26 $i^\prime$-band magnitude, but this can be attributed to spurious detections around image edges, stellar diffraction, and noise spikes. Table \ref{tab:5sig} breaks down this information by the Subaru footprint for the optical and near-IR data.

\subsection{Near-Infrared Observations}
The SuperCLASS field was observed for 12.7 hours with {\it Spitzer} {\sc IRAC} at $3.6\,\micron$ ($M$-band) and $4.5\,\micron$ ($N$-band) during Cycle 12 in program \#12074 (PI Casey). Flux calibration was completed using astronomical standard stars in {\it Spitzer}'s continuous viewing zone and that were monitored throughout observations. Data was reduced and processed with the Basic Calibrated Data (BCD) pipeline and post-BCD software. Sources were extracted using {\tt SExtractor} accounting for the large point response function (PRF)
, 1.78\arcsec\ for the 3.6\,\micron\ channel and 1.80\arcsec\ for the 4.5\,\micron\ channel. No prior positions were used for {\it Spitzer} source extraction and detections were treated as independent from sources identified in the optical bands. Imaging reached a 5\,$\sigma$ point source sensitivity limit of 21.1 magnitudes at both $3.6\,\micron$ and $4.5\,\micron$. Catalogs are approximated to be 90\% complete down to 20.0 and 20.1, respectively, finding $\approx$82,000 and $\approx$83,000 sources in the two filters across the survey. Source positions are matched to the optical catalog with a nearest neighbor radius of $1\arcsec$. The spherical distance to the nearest optical sources is 0.3\arcsec\ and after injecting 80,000 fake sources into the {\it Spitzer} field we find a false match contamination rate of 1.25\%. 

Near-IR $K$-band observations of the SuperCLASS field were obtained with WIRCAM on the Canada-France-Hawaii Telescope (PI Chapman). Images reached moderate depth (AB\,$<$\,19), but unfortunately a key section of the field (encompassing the DR1 \eMERLIN\ field) lacks coverage. Because of this gap in observations and relatively shallow and non-uniform depth, $K$-band photometry is not included in the photometric redshift analysis outlined in Section \ref{sec:phot-z}.

\begin{table}
\begin{center}
\caption{Optical/NIR Completeness}
\label{tab:5sig}
\begin{tabular}{cccc}
\hline
Footprint & Filter & 5$\sigma$ Detection & 90\% \\
 & & Threshold & Completeness \\
\hline
1 & B & 26.1 & 24.9 \\ 
1 & V & 26.1 & 24.9 \\
1 & r$^\prime$ & 26.0 & 24.8 \\
1 & i$^\prime$ & 25.7 & 24.6 \\
1 & z$^\prime$ & 24.9 & 24.6 \\
\hline
2 & B & 26.1 & 24.8 \\ 
2 & V & 25.5 & 24.7 \\
2 & r$^\prime$ & 25.6 & 24.6 \\
2 & i$^\prime$ & 25.2 & 24.7 \\
2 & z$^\prime$ & 25.0 & 24.5 \\
\hline
3 & B & 26.1 & 24.8 \\ 
3 & V & 26.0 & 24.7 \\
3 & r$^\prime$ & 26.2 & 24.5 \\
3 & i$^\prime$ & 25.9 & 24.5 \\
3 & z$^\prime$ & 25.4 & 24.4 \\
\hline
4 & B & 26.3 & 24.9 \\ 
4 & V & 26.1 & 24.6 \\
4 & r$^\prime$ & 25.9 & 24.5 \\
4 & i$^\prime$ & 25.8 & 24.5 \\
\hline
5 & B & 26.3 & 24.8 \\ 
5 & V & 26.3 & 24.3 \\
5 & r$^\prime$ & 25.2 & 24.6 \\
5 & i$^\prime$ & 25.2 & 24.6 \\
\hline
6 & B & 26.3 & 24.8 \\ 
6 & V & 26.1 & 24.6 \\
6 & r$^\prime$ & 25.2 & 24.6 \\
6 & i$^\prime$ & 25.2 & 24.6 \\
\hline
full field & Y & 25.9 & 23.8 \\ 
full field & [3.6] & 21.1 & 20.0 \\
full field & [4.5] & 21.1 & 20.1 \\
\hline
\end{tabular}
\end{center}

{\bf Note --} List of 5$\sigma$ detection thresholds and 90\% completeness limits for each photometric band, sorted by footprint. {\it z}$^{\prime}$-band data is currently only available for footprints 1-3.
\end{table}

\subsection{L-band and Other Radio Observations}
The \eMERLIN\, observations consist of 112 pointings, of which 49 have been observed, making up the the current 0.26\,deg$^2$ Data Release 1 (DR1) area. Each pointing was observed for 400 minutes in total 
with an RMS sensitivity of 15\,\uJy\, at 1.4\,GHz. This region includes four of the five Abell clusters (A968, A981, A998, and A1005). Observations were performed in units of 15 hours over the frequency range 1.204--1.717\,GHz. The bandwidth was divided into 8 IFs, each of 512 channels of 125 kHz width to give an unsmeared field of view (FOV) of 30\arcmin, sufficient to map the whole primary beam of the smallest (25\,m diameter) telescopes in the \eMERLIN\ array. Observations began 2014 December 29, continuing into 2015 and 2016. Data were reduced with a modified version of the NRAO AIPS package and cleaned of radio frequency interference (RFI) using a similar algorithm to the AOFlagger program \citep{Offringa2010,Offringa2010a}. In Figure \ref{fig:SC_map}, the dark gray area shows the completed observations of the DR1 region and white encompasses the full field as well as the proposed extension to the south which would increase the survey area to $\sim\,$2\,deg$^2$ and include the fifth cluster. Please refer to Paper I in this series for more detailed discussion of the first radio data release. 

In addition, 24 hours of L-band (1.5\,GHz) observations (112 pointings) were taken using the VLA in A configuration in August 2015 under project code 15A-053 (PI Battye) to complement the \eMERLIN\ observations. The total observing time was divided into 6$\times$4 hour sessions and were assigned staggered LST start times to maximize coverage in the $u-v$ plane. These observations yielded an RMS noise typically less than 7\,\uJy\ beam$^{-1}$ across the mosaic.

Low-frequency 325\,MHz observations were carried out using the Giant Metrewave Radio Telescope (GMRT) (PI Scaife) covering a $\sim\,$6.5\,deg$^{2}$ FOV, achieving a nominal sensitivity of 34\,\uJy\ beam$^{-1}$. From these data, a catalog of 3257 sources with flux densities in the range of 183\,\uJy--1.5\,Jy was compiled \citep{Riseley2016}; 454 of these sources lie in the DR1 field.

\section{Photometric Redshifts}
\label{sec:phot-z}
\subsection{Photometric Redshift Fitting}
\label{subsec:phot-z_fitting}
We use the Easy and Accurate Z-phot from Yale \citep*[{\sc EAZY};][]{Brammer2008} to calculate photometric redshifts across the SuperCLASS field. {\sc EAZY} is optimized for galaxy samples lacking spectroscopic information with the default settings optimized to work with multiwavelength photometric data alone, though spectroscopic redshifts can be included for photometric redshift calibration if available. {\sc EAZY} also includes dusty starburst models as part of its template set, an important addition considering our interest in DSFGs.

\begin{figure*}
\begin{minipage}[t]{0.3\textwidth}
\includegraphics[width=\textwidth]{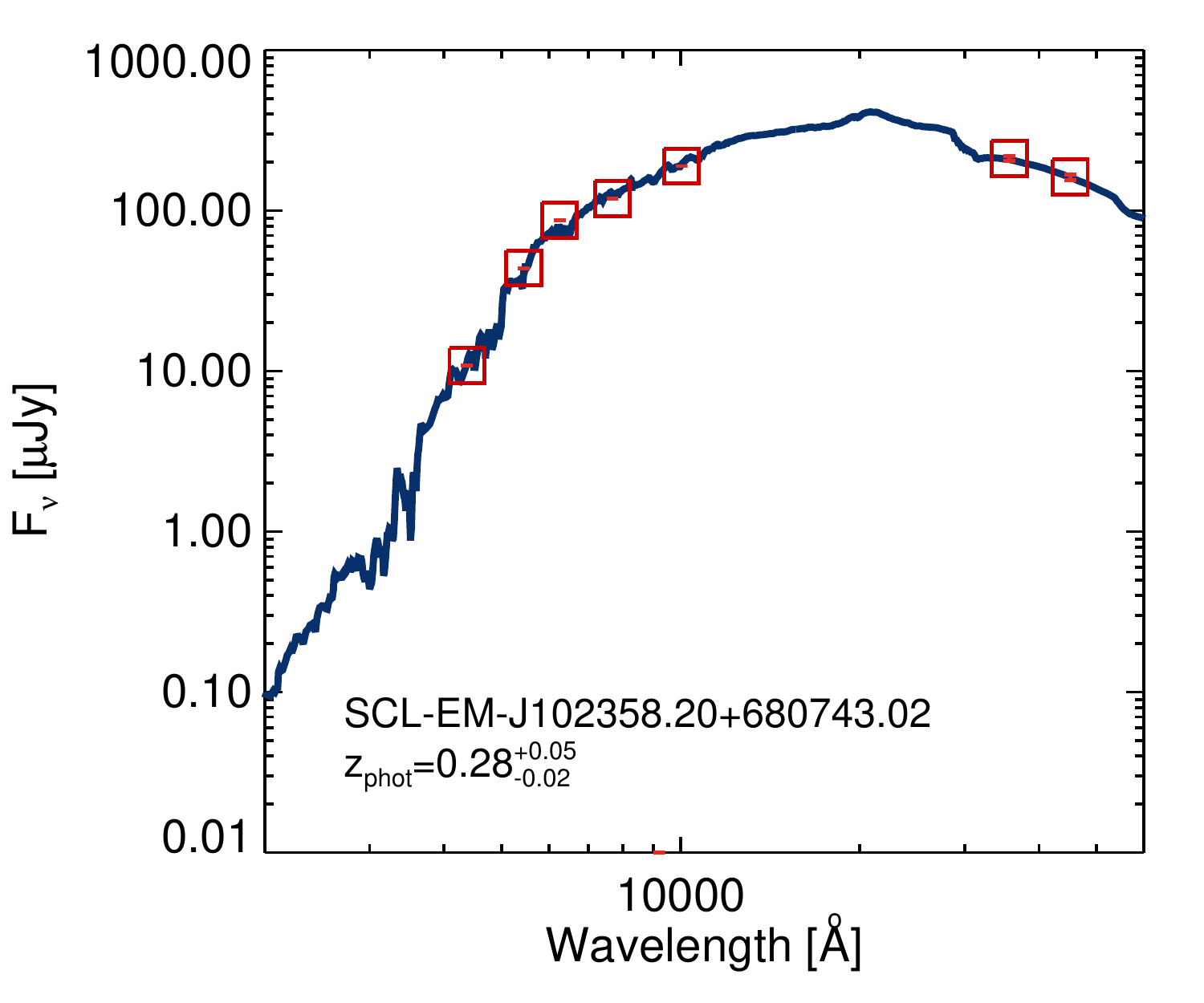}
\end{minipage}
\begin{minipage}[t]{0.3\textwidth}
\includegraphics[width=\textwidth]{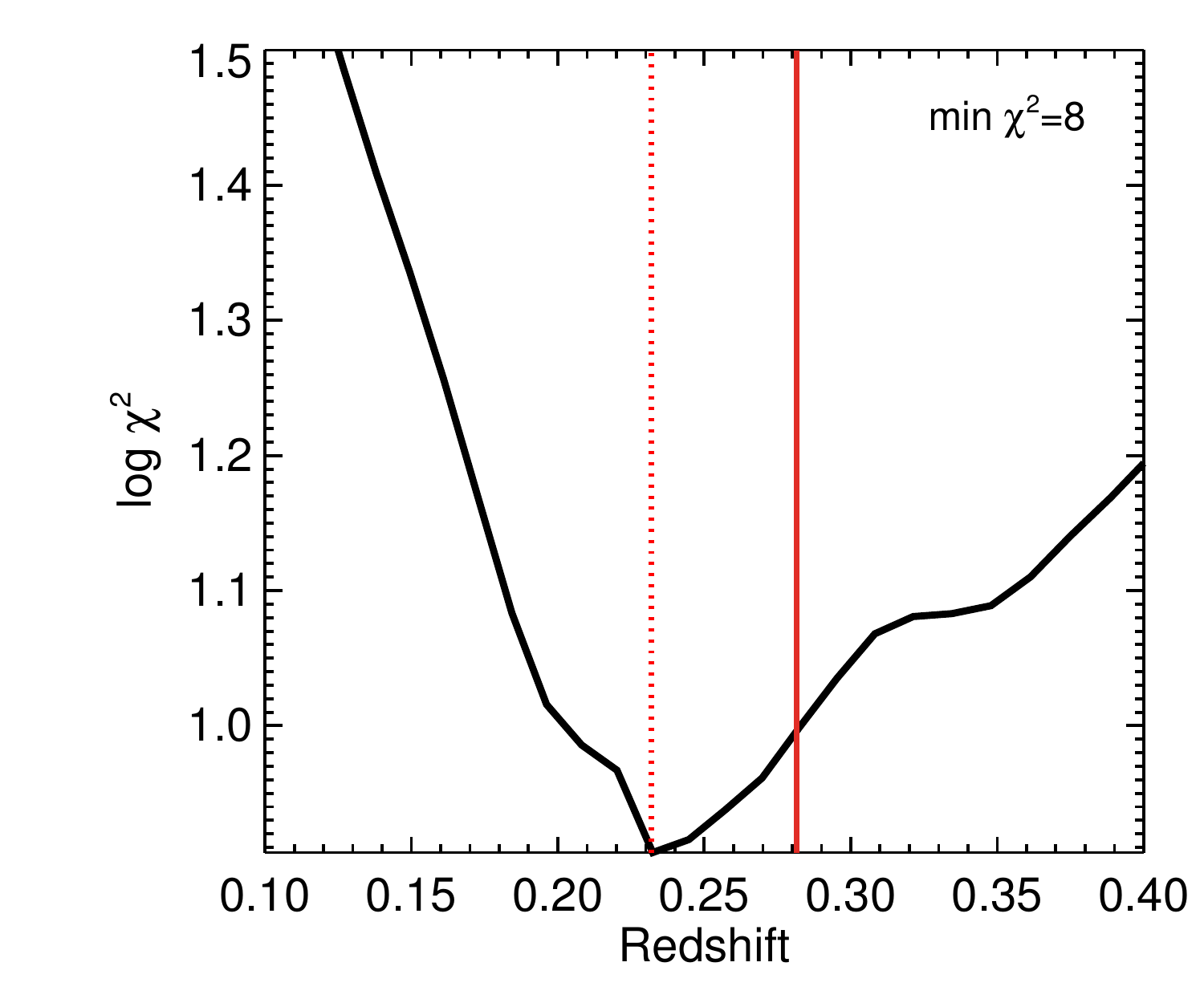}
\end{minipage}
\begin{minipage}[t]{0.3\textwidth}
\includegraphics[width=\textwidth]{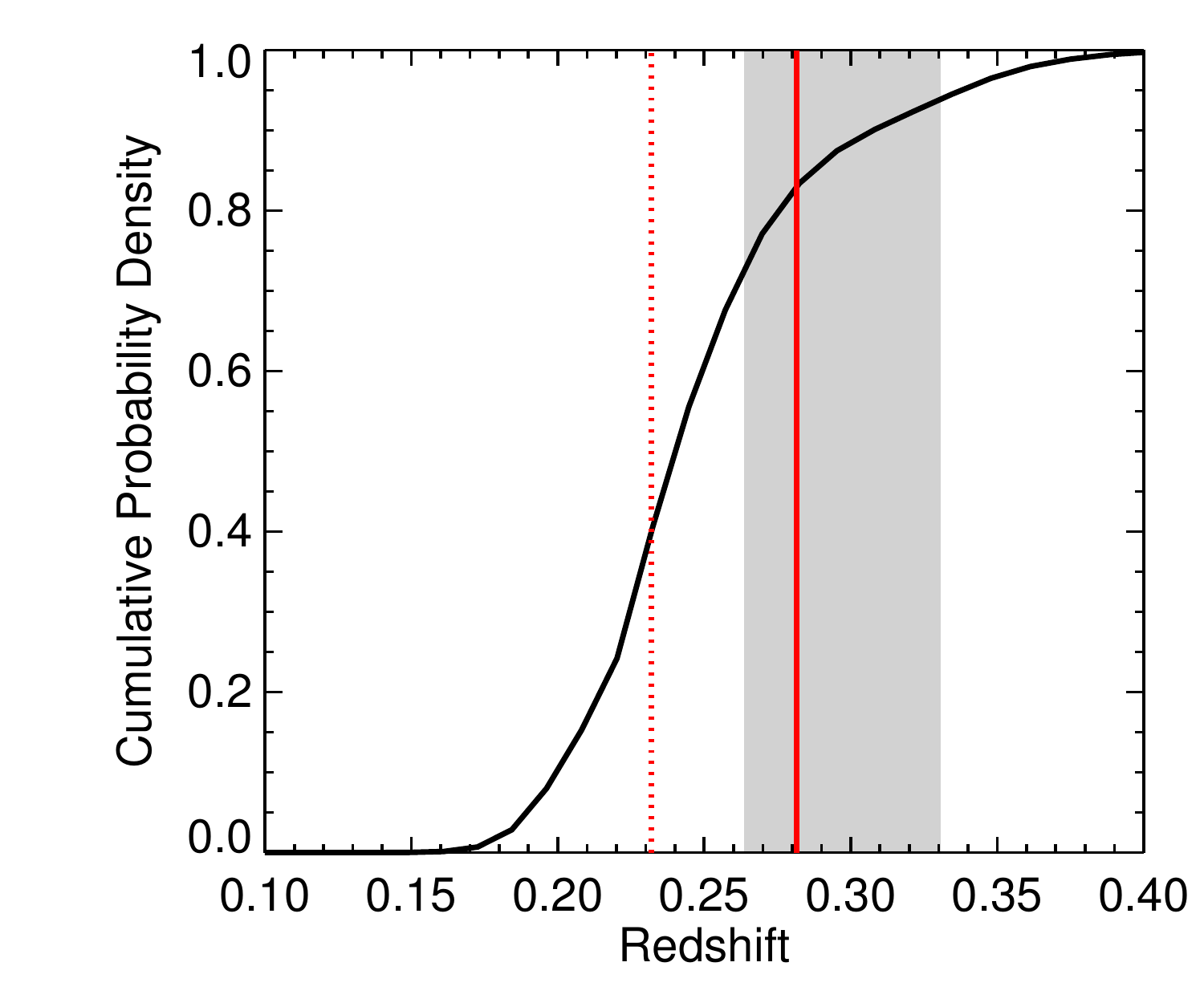}
\end{minipage}
\begin{minipage}[t]{0.3\textwidth}
\includegraphics[width=\textwidth]{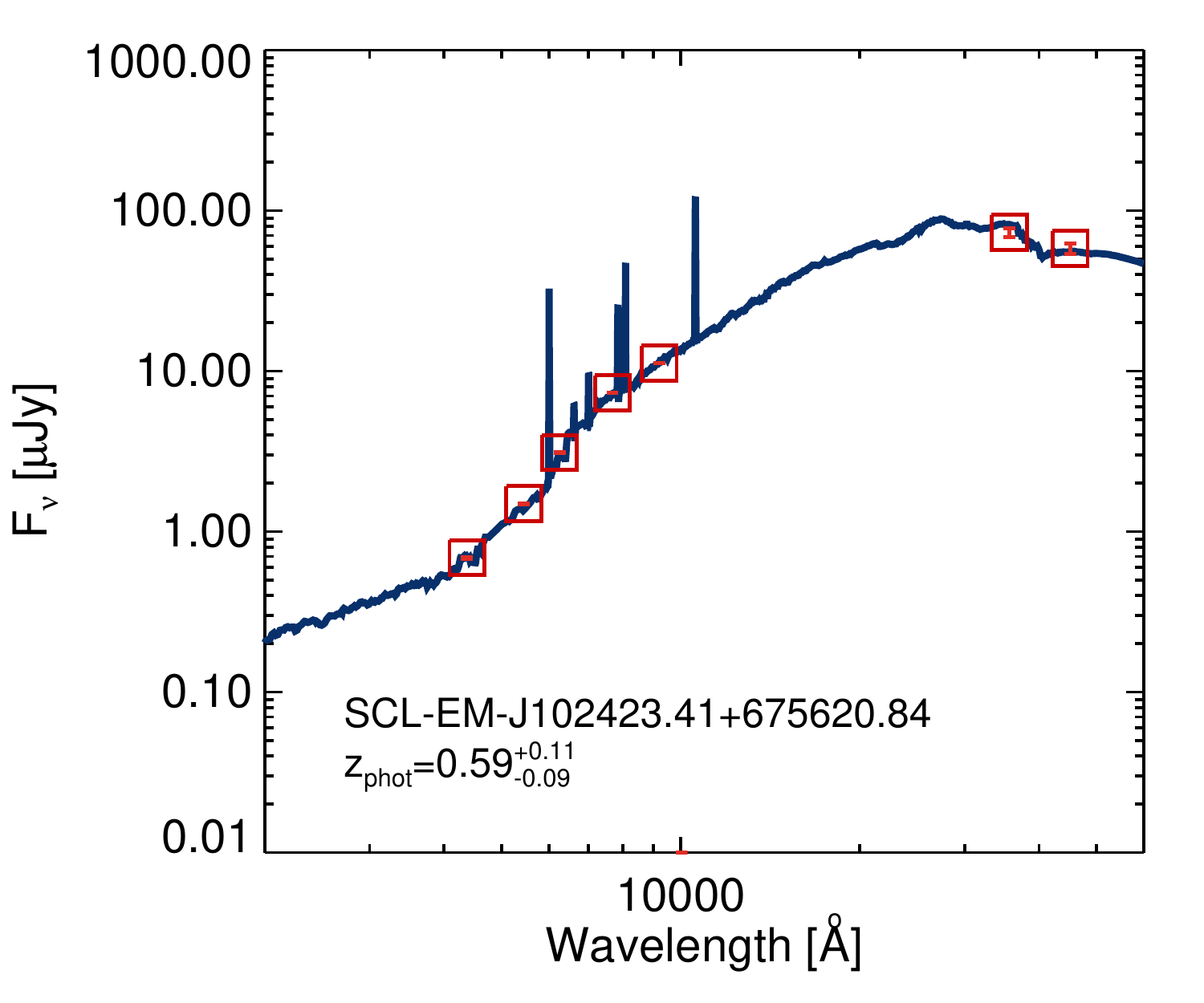}
\end{minipage}
\begin{minipage}[t]{0.3\textwidth}
\includegraphics[width=\textwidth]{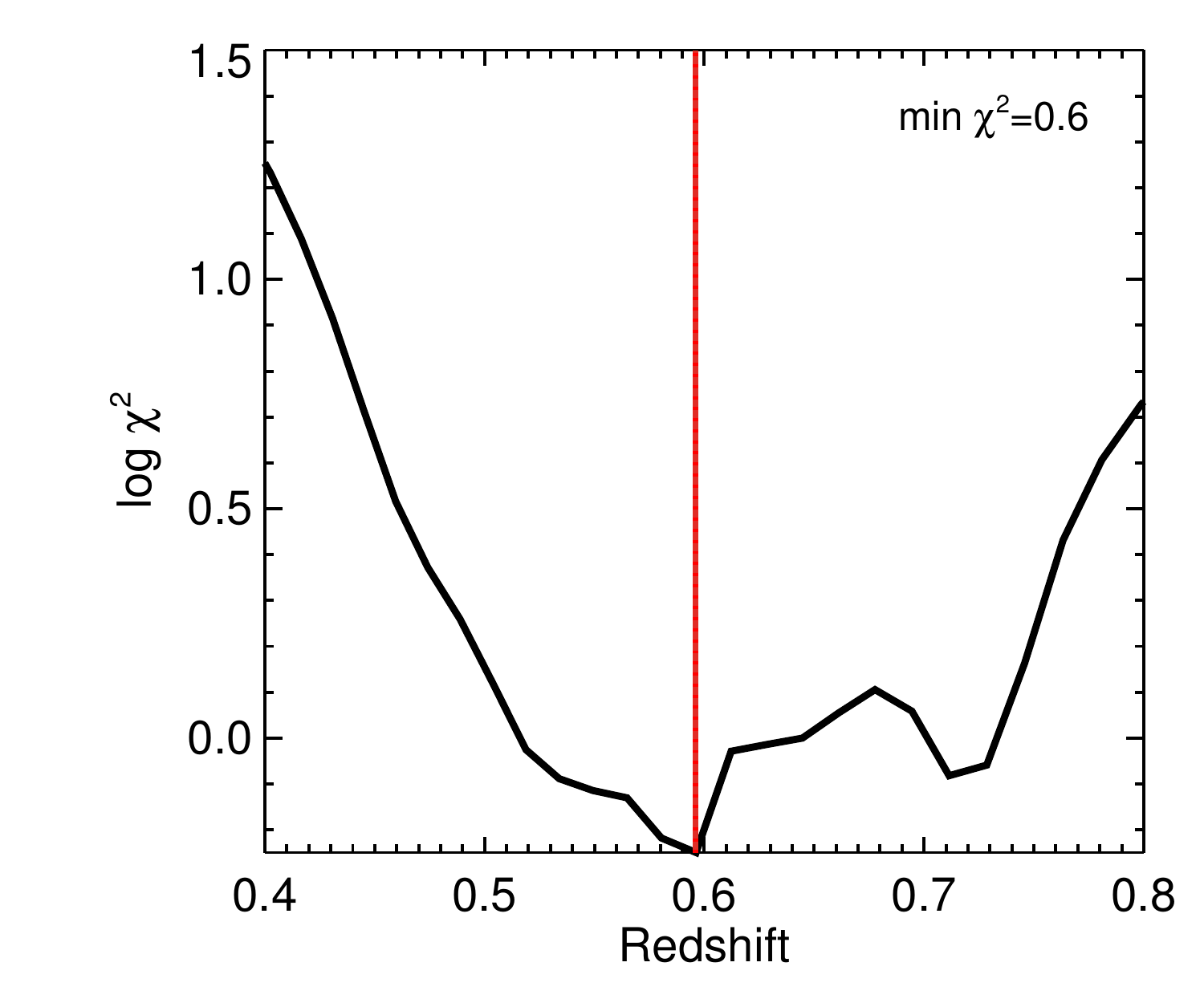}
\end{minipage}
\begin{minipage}[t]{0.3\textwidth}
\includegraphics[width=\textwidth]{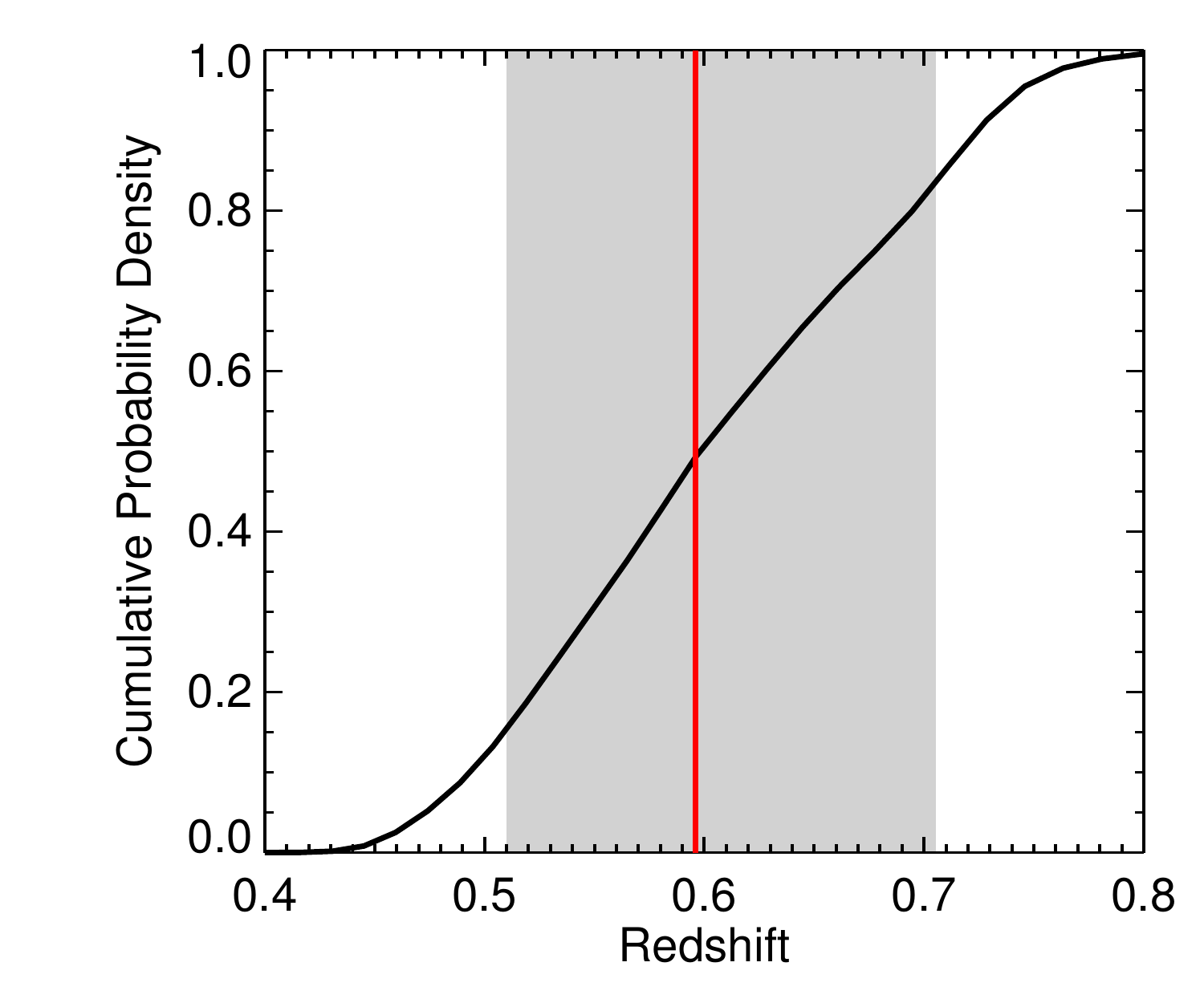}
\end{minipage}
\begin{minipage}[t]{0.3\textwidth}
\includegraphics[width=\textwidth]{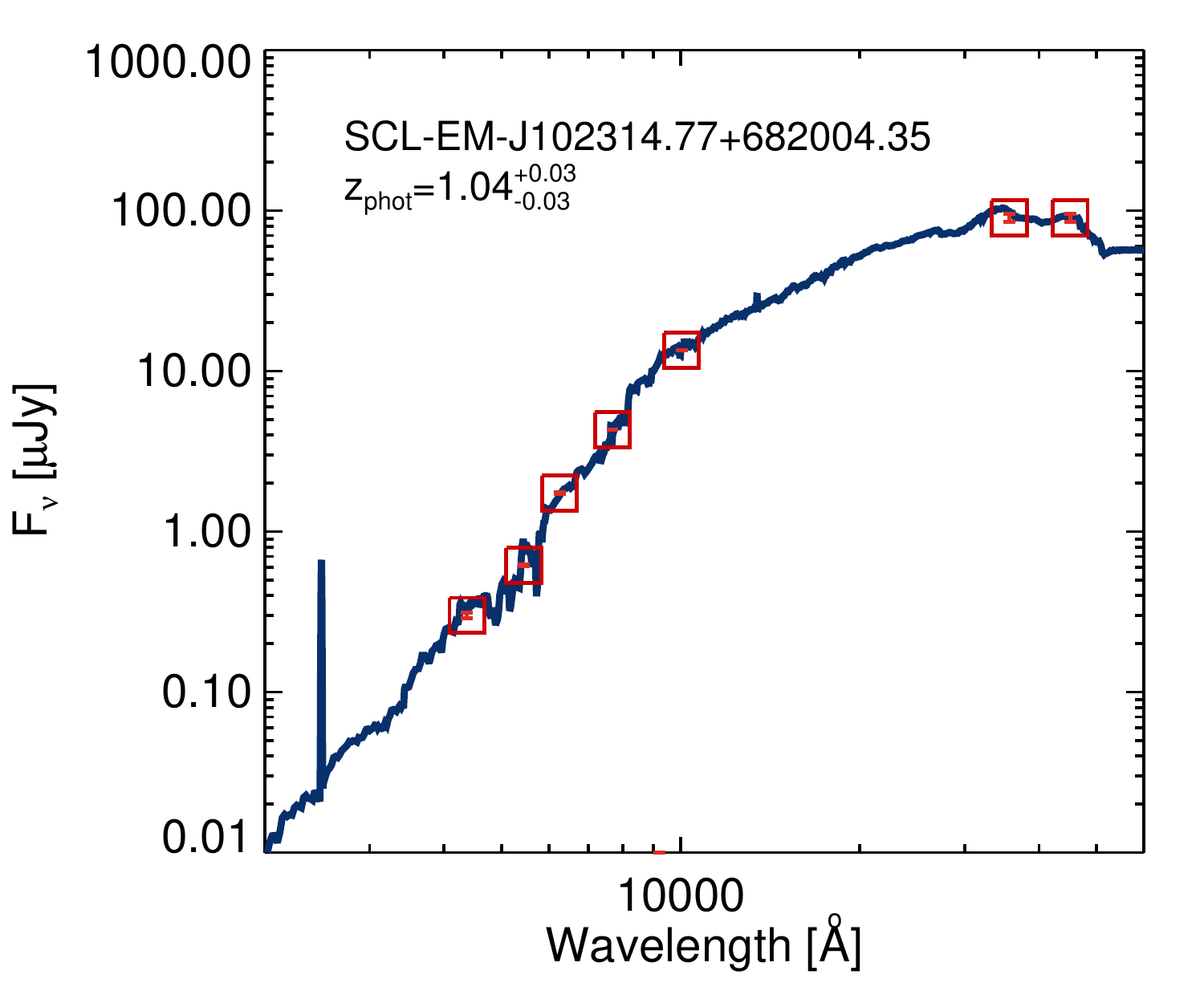}
\end{minipage}
\begin{minipage}[t]{0.3\textwidth}
\includegraphics[width=\textwidth]{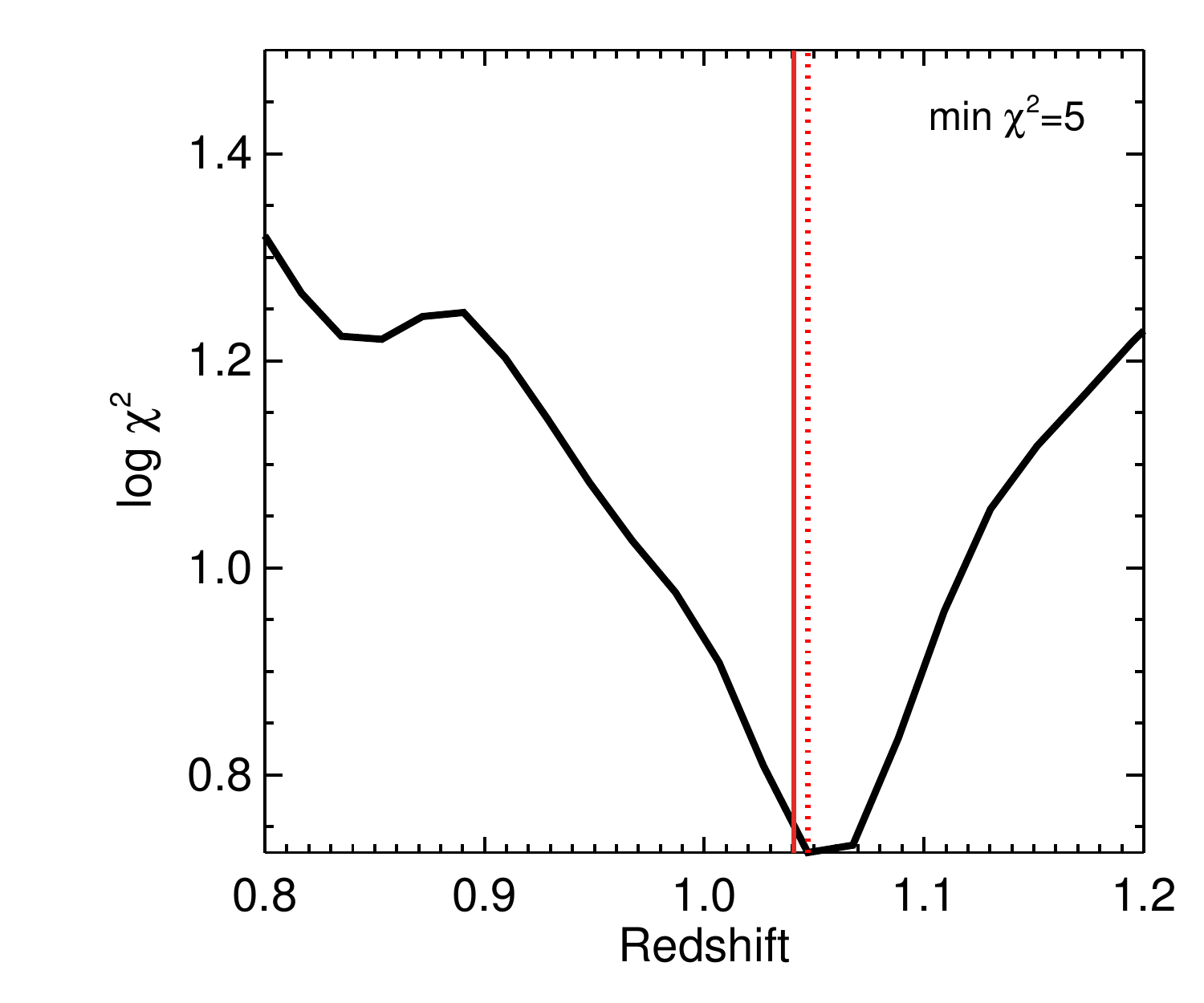}
\end{minipage}
\begin{minipage}[t]{0.3\textwidth}
\includegraphics[width=\textwidth]{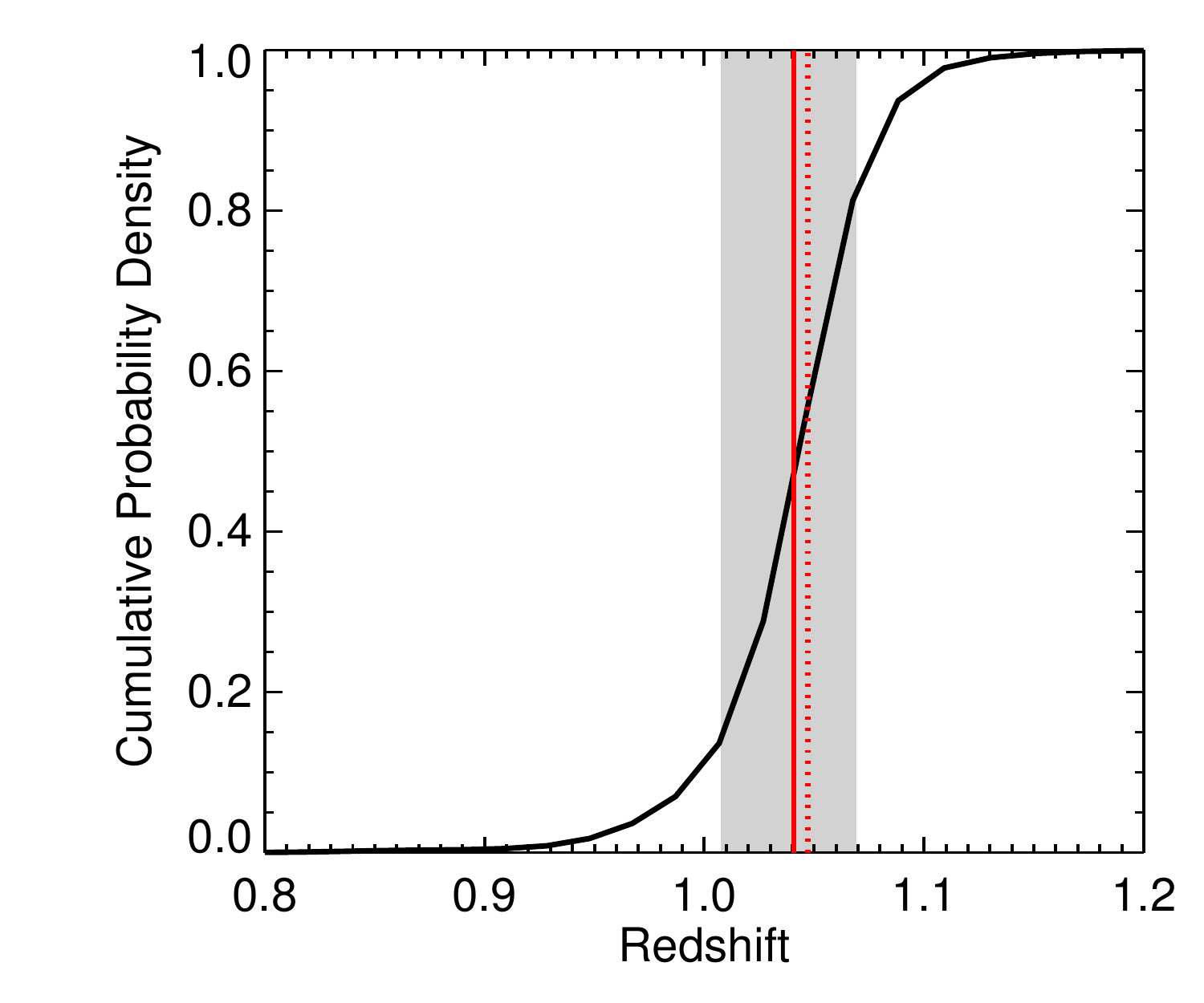}
\end{minipage}
\begin{minipage}[t]{0.3\textwidth}
\includegraphics[width=\textwidth]{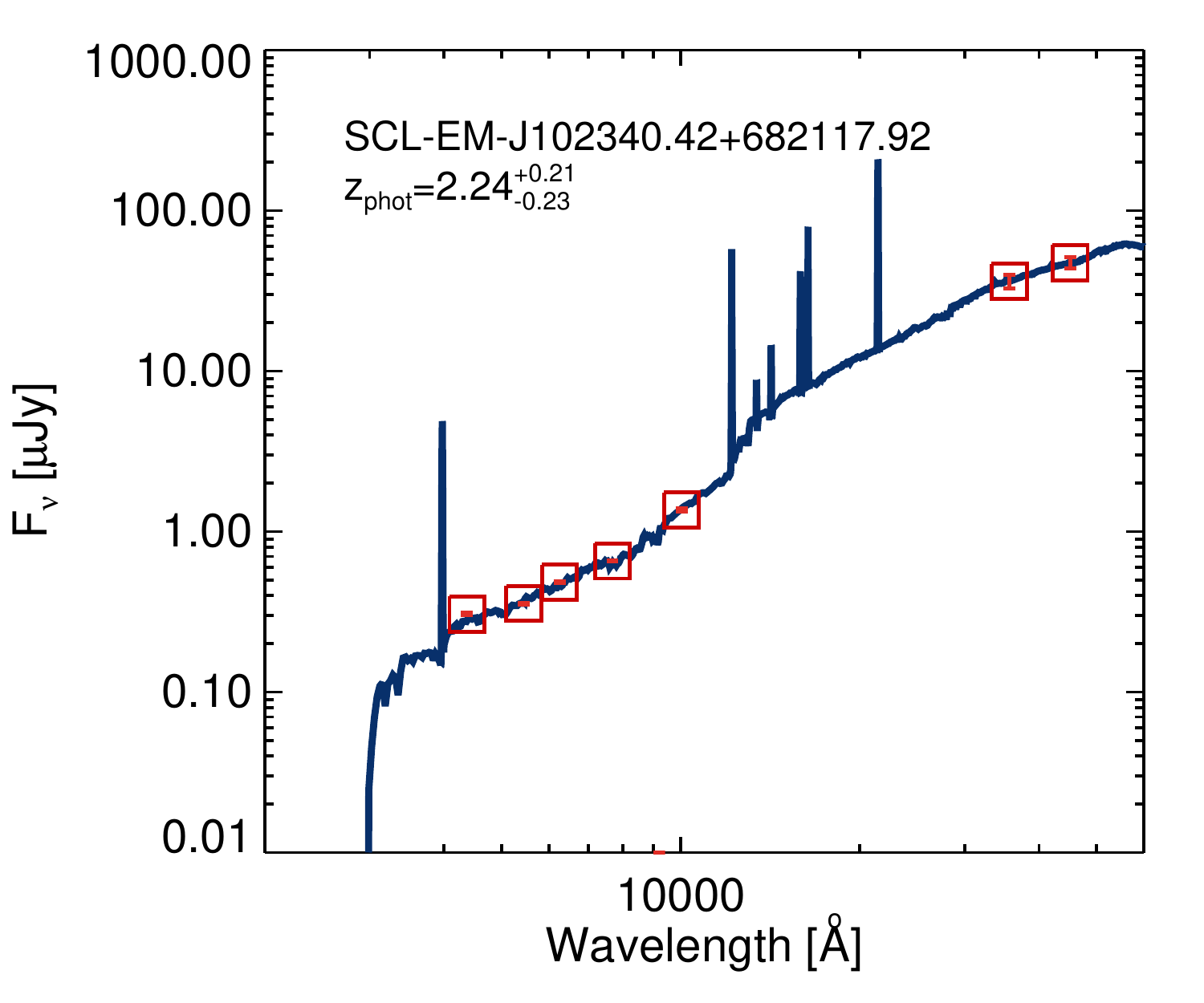}
\end{minipage}
\begin{minipage}[t]{0.3\textwidth}
\includegraphics[width=\textwidth]{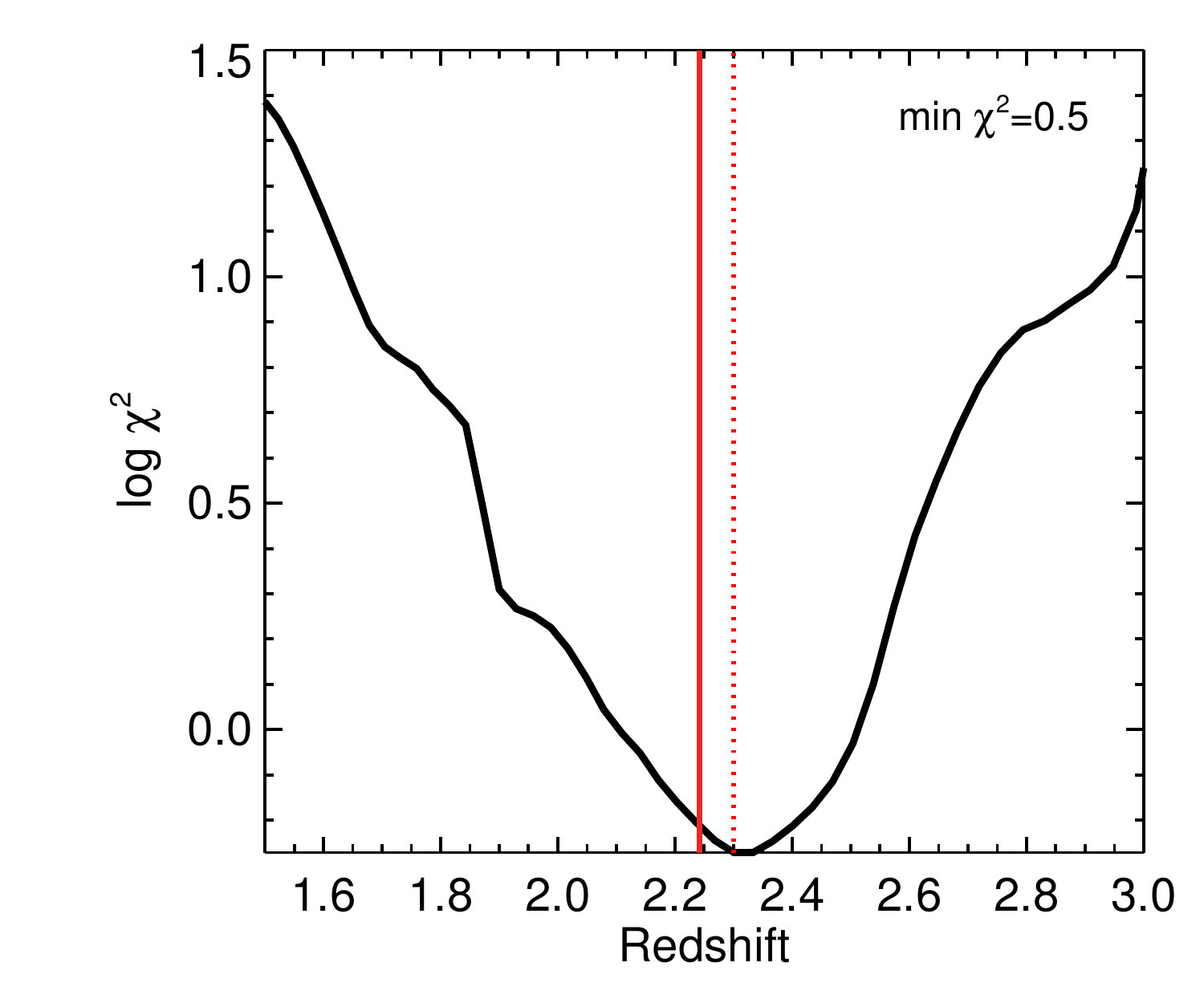}
\end{minipage}
\begin{minipage}[t]{0.3\textwidth}
\includegraphics[width=\textwidth]{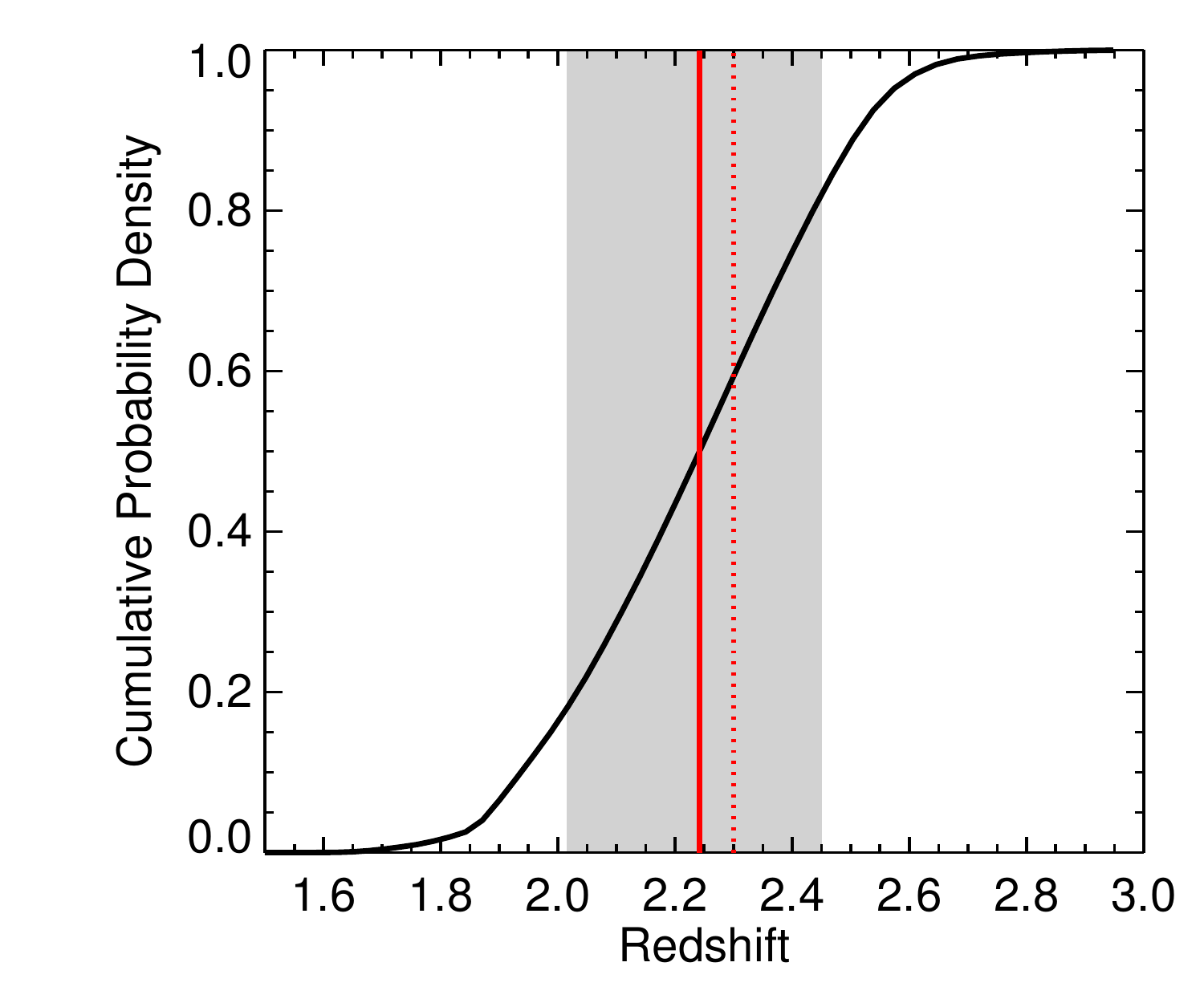}
\end{minipage}
\begin{minipage}[t]{0.3\textwidth}
\includegraphics[width=\textwidth]{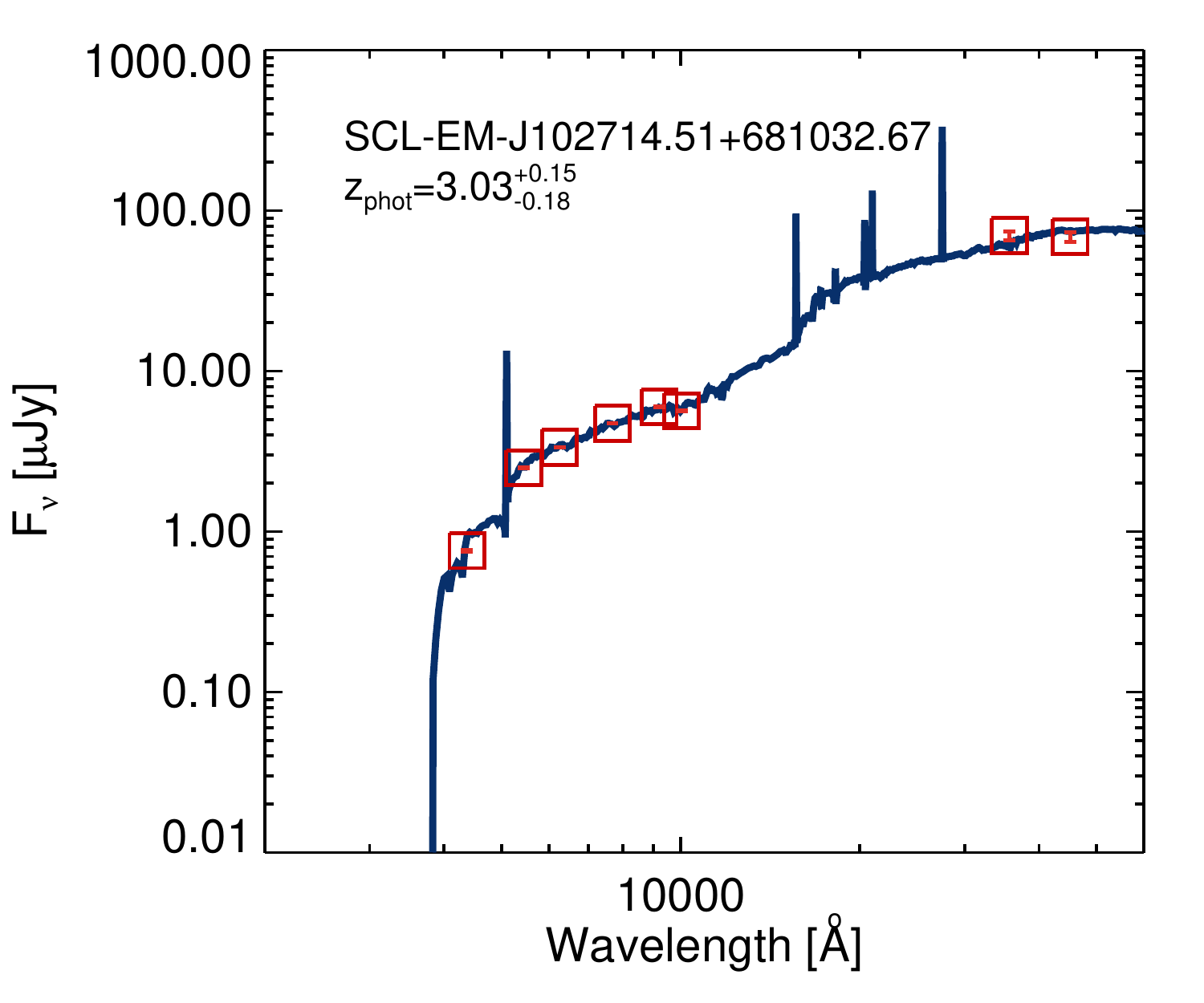}
\end{minipage}
\begin{minipage}[t]{0.3\textwidth}
\includegraphics[width=\textwidth]{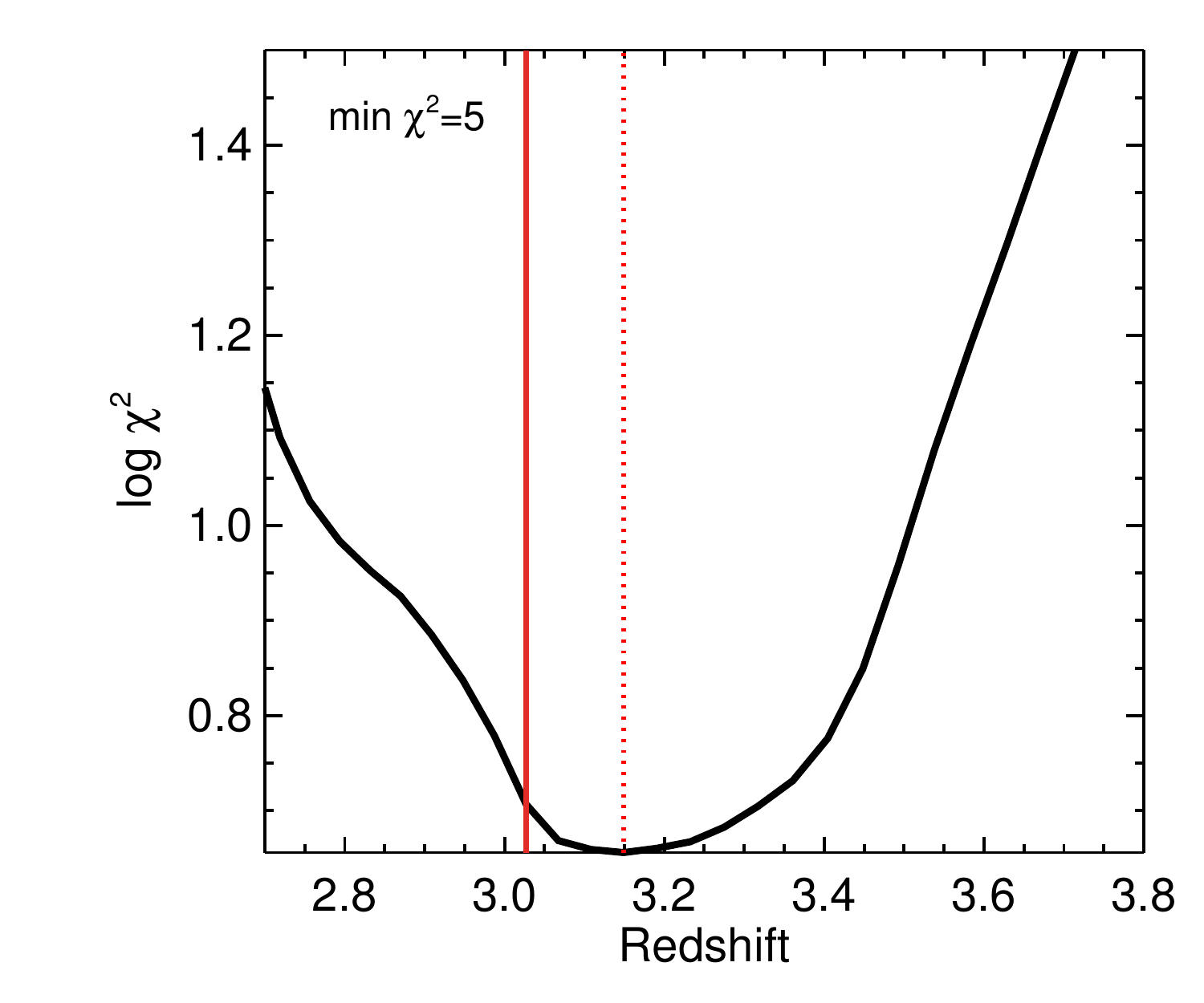}
\end{minipage}
\begin{minipage}[t]{0.3\textwidth}
\includegraphics[width=\textwidth]{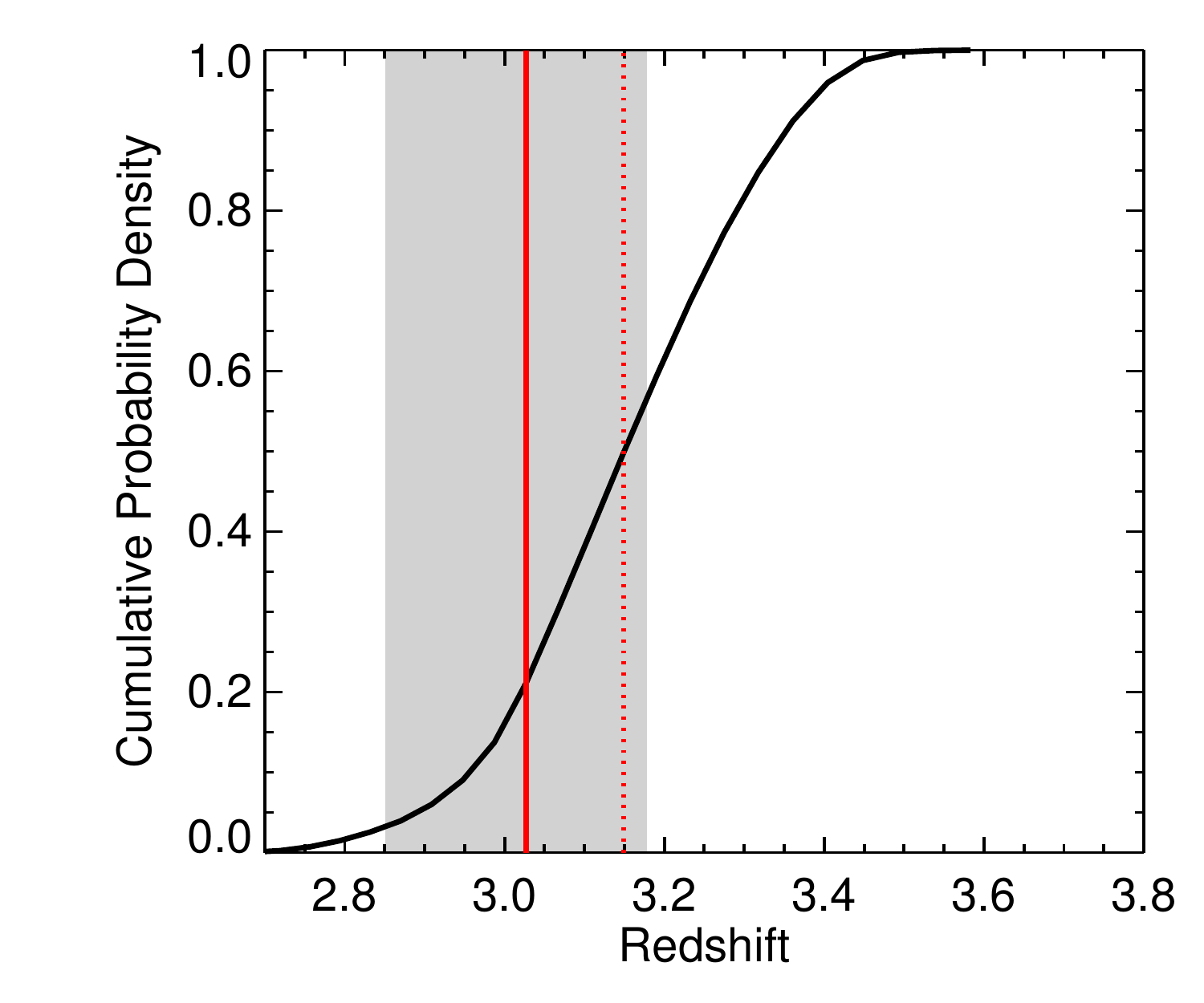}
\end{minipage}
\caption{Example spectral energy distributions (SED) fit to SuperCLASS galaxy photometry (\textit{left}), $\chi^2$ distribution of the best fit star-forming template sets (\textit{center}), and cumulative probability density function (\textit{right}), for five sources with $0<z_{\rm phot}<4$. Dotted red lines show the photometric redshift solution corresponding to the minimum $\chi^2$. Solid red lines mark photometric redshifts after the COSMOS priors are applied and their corresponding 1\,$\sigma$ confidence intervals in gray.}
\label{fig:ez_outs}
\end{figure*}

The template set provided by {\sc EAZY} (v1.3 which includes a dusty starburst model) is a linear combination of $\sim\,$500 \cite{Bruzual2003} (BC03) synthetic galaxy photometry models which are then used to create eight ``basis'' templates. These eight templates are then combined using weighting coefficients to create a representative sample of galaxy spectral energy distributions (SEDs) over a broad redshift range. The algorithm steps through a grid of redshifts ($0<z<6$) and at each redshift finds the best-fitting synthetic template spectrum. Using {\sc EAZY} we obtain the best-fit SED generated from the template set, the photometric redshift probability distribution, and the minimum $\chi^{2}$ distribution as a function of redshift for the best fit template for each source. We show five example galaxies with good photometric redshift fits ranging from $0.2<z<3.1$ in Figure \ref{fig:ez_outs}.

\citet{Duncan2018} showed that some improvement in the photometric redshift estimation for radio sources is possible using alternative template sets within {\sc EAZY}, particularly the ATLAS \citep{Brown2014} and {\it XMM}-COSMOS \citep{Salvato2009,Salvato2011} templates. The current {\sc EAZY} template set only models stellar emission and does not include any contribution from AGN (e.g., contribution from AGN continuum  emission or broad emission lines), which could result in inaccurate redshift fits for our radio source population. Both the radio source population and the potential AGNs are of interest to us for both accurate redshifts for weak lensing and galaxy evolution analysis. Thus, we also use the {\it XMM}-COSMOS template set to run an independent test of photometric redshifts for all sources for comparison.

\begin{figure}
\includegraphics[width=\columnwidth]{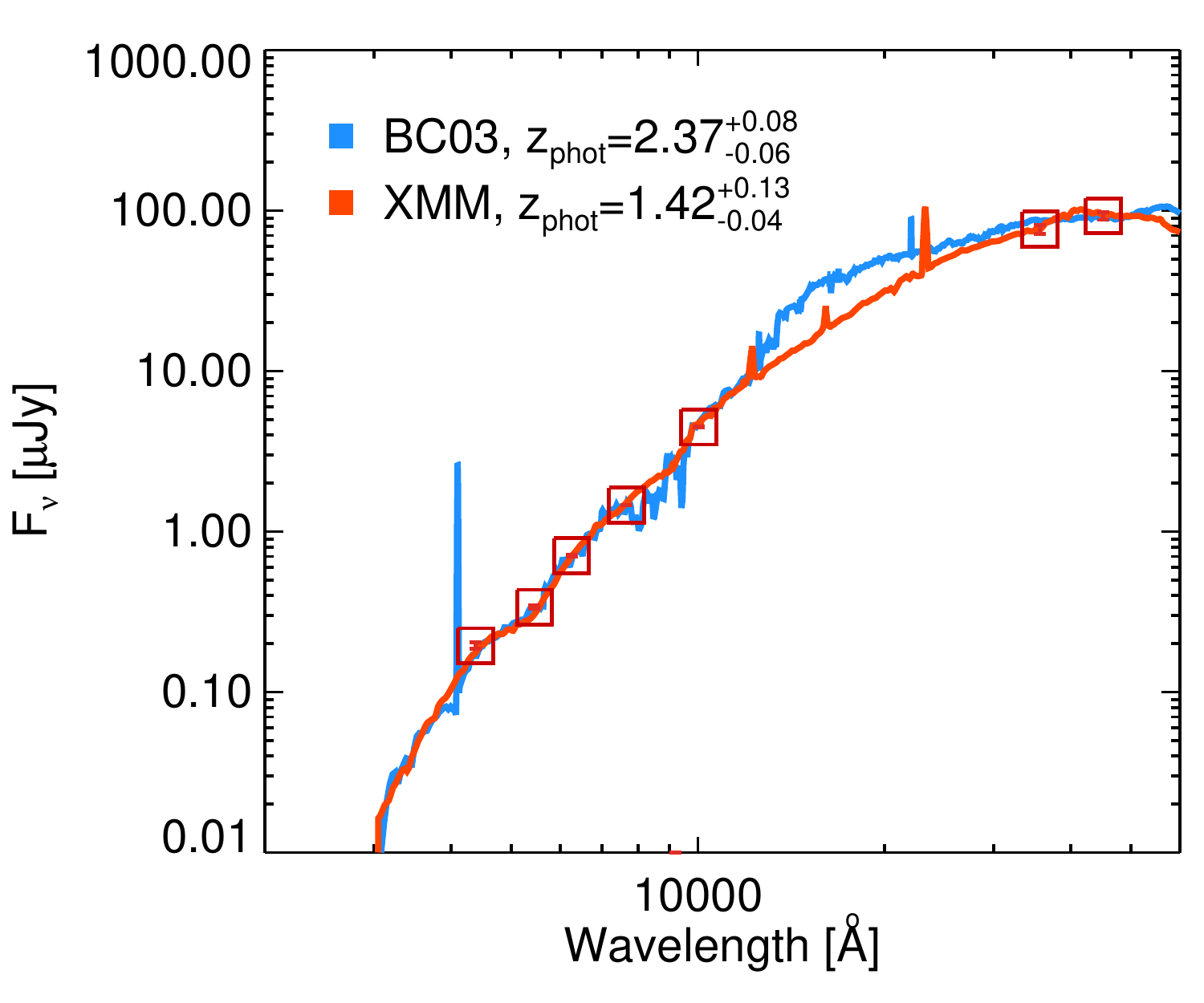}
\vspace{-5mm}
\caption{An example of an SED generated by {\sc EAZY} using star-forming (blue) and AGN (orange) template sets illustrating a degeneracy in $z_{\rm phot}$ solution which occurs for a handful of sources with similar red spectral shapes. The lack of distinctive breaks due to limited photometry make unambiguous identification difficult.}
\label{fig:sed_compare}
\end{figure}

In the central DR1 SuperCLASS map 154 \eMERLIN/VLA \uJy\ radio sources are identified. Of these sources, 149 ($\sim\,$97$\%$) have good or moderate quality photometric redshift solutions with $\Delta z_{\rm phot}/(1+z_{\rm phot})<1$, where $\Delta z_{\rm phot}=z_{\rm u68}-z_{\rm l68}$. We opt against using a stronger cut because of the limited number of filters.
The 5 remaining radio sources have highly degenerate photometric redshift solutions spanning a wide redshift range. For the 149 sources with converged photometric redshift fits, we visually inspect the SEDs produced from the star-forming templates (BC03) and AGN templates ({\it XMM}) to determine the best-fit solution. We refrain from using minimum $\chi^2$ values to compare the fits using BC03 or {\it XMM} due to the different number of free parameters in their respective photometric redshift determinations. A few cases do not have substantially better fits in either the BC03 or {\it XMM} template set. While the majority of these exhibit similar redshift solutions, a few do not. An example of one such source is illustrated in Figure \ref{fig:sed_compare}. In cases such as this, we defer to the BC03 fit given the range of radio flux densities in our sample and the lower anticipated source density of AGN, but keep track of these cases and a future paper will further explore their photometry and morphologies to infer a more accurate constraint.

The redshift distribution for all $176,000+$ OIR sources in the $\sim\,$1.53\,deg$^2$ field appears to show a deficit of sources between $1<z_{\rm phot}<2$ and an excess at $z_{\rm phot}<1$ relative to fields with well-characterized photometric redshifts \citep*[e.g. COSMOS;][]{Laigle2016}. To determine the origin of these discrepancies, we employ a mock catalog of galaxies from cosmological simulations to examine the effectiveness of our filter set in recovering photometric redshifts (Section \ref{sec:mocks}). 

\subsection{Assessing Photometric Redshift Errors via Close Pairs}
\label{subsec:close_pairs}
Following the work of \citet{Quadri2010} and \citet{2016ApJ...822...30B}, we determine the scatter in our photometric redshifts from an analysis of close pairs to provide confidence in the precision of the estimates presented in this paper. As stated in these works, galaxies which appear near to each other (2.5--15\arcsec separation) are likely to be physically associated thanks to the clustered nature of galaxies. If we assume both galaxies in a projected pair are at the same redshift, then any discrepancy between the two measured redshifts will be due to an error in the measurements and fitting procedure. This close pairs analysis allows for an independent test of the photometric redshift accuracy which is necessary considering the lack of spectroscopic redshifts available for comparison in our sample.

Errors on the photometric redshifts of the close pairs are related to the Gaussian distribution of their redshift deviations \citep[Equation 2 of ][]{2016ApJ...822...30B} and we adopt a separation range of 2.5--15\arcsec to determine pairs. From the distribution of redshift separations for galaxy pairs, we then subtract the distribution created using a catalog in which the galaxy positions have been randomized in order to estimate the contribution from chance projections. After fitting a Gaussian to the new distribution accounting for uncorrelated pairs, we determine the representative photometric redshift scatter, $\sigma_{\rm pairs}/\sqrt{2}=0.11$, to be in line with our measured uncertainties which are 0.08 on average.

\subsection{Photometric Redshift Priors With COSMOS}
\label{subsec:cosmos_priors}
Based on the assumption that the \uJy\ radio source redshift distribution is similar across all areas of the sky, we use the well-constrained photometric redshifts for 1.4\,GHz radio galaxies from the COSMOS field \citep{Laigle2016} as priors to weight the probability distribution functions (PDFs) generated for the 154 \eMERLIN\ detected radio sources by {\sc EAZY}. For each source, we find all sources in the COSMOS catalog that have similar VLA 1.4\,GHz flux densities and 3.6\,\micron\ magnitudes (within 5\%). We then construct a redshift distribution, taking the individual PDFs from COSMOS photometry into account, and multiply it by the {\sc EAZY} PDF to obtain a redshift distribution based on the prior as matched to the COSMOS sample. Six out of 149 radio sources do not have 3.6\,\micron\ detections and as such do not have a prior applied to their photometric redshift solution. The photometric redshifts of the radio sources reported throughout this paper are the median of the distribution produced by the product of the original {\sc EAZY} PDFs and the prior generated from COSMOS.

\section{Mock Catalog from Cosmological Simulations}
\label{sec:mocks}
As a confidence check of our work, we run a mock catalog of sources, described in \citet{Jouvel2009}, through {\sc EAZY} to understand the validity of {\sc EAZY} estimated redshifts for given photometric information. The mock catalog has the same error properties as the OIR data and with it we test the reliability of the photometric redshifts and how improvements can be made, specifically with the introduction of more photometric bands.

\begin{figure}
\includegraphics[width=\columnwidth]{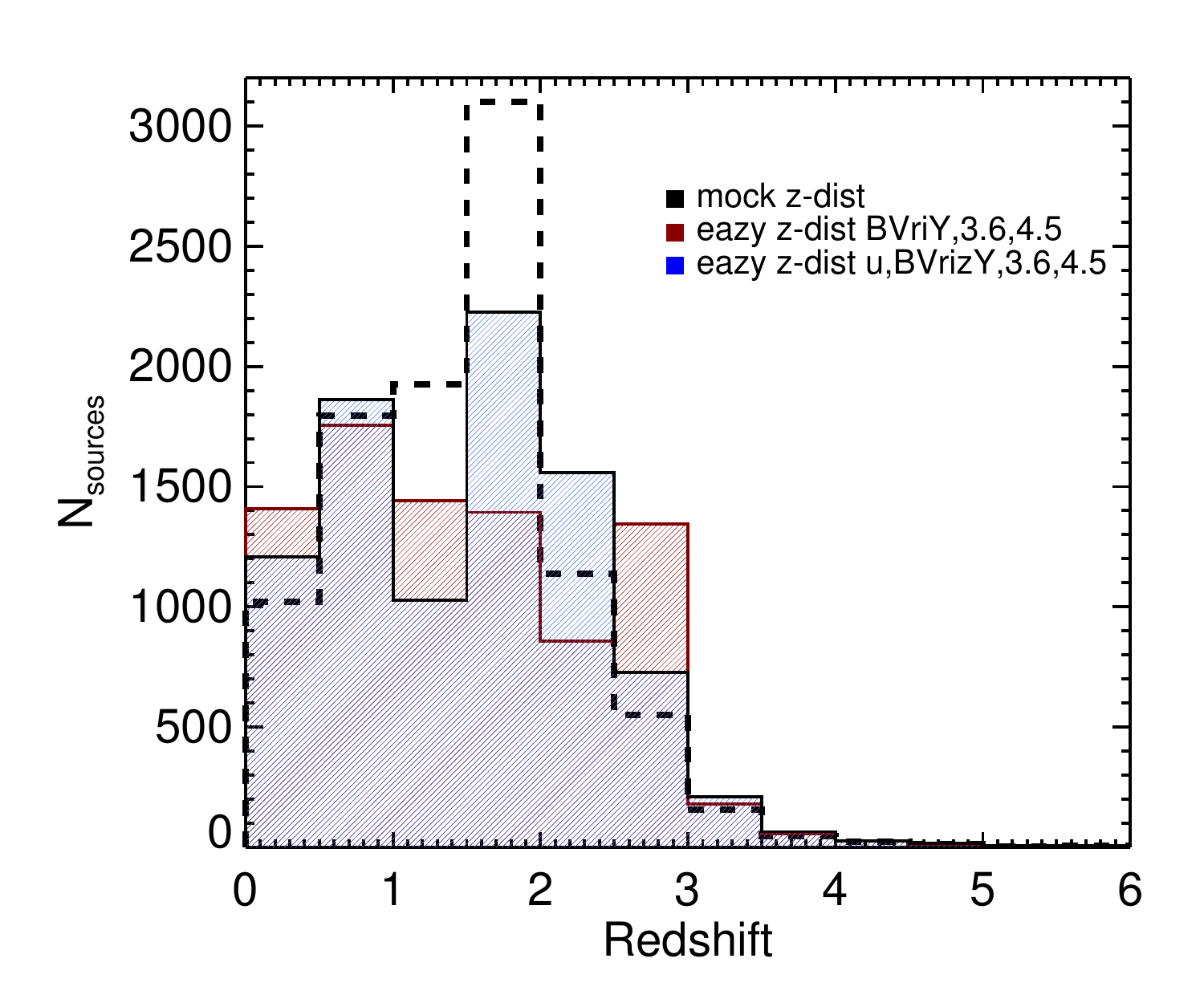}
\vspace*{-5mm}
\caption{Comparison on the redshift distributions of mock catalogs with and without \textit{u} and \textit{z}$^\prime$-band data. These two filters are particularly important for picking up the Lyman and Balmer breaks, significant spectral features that inform photometric redshift fitting, at $z<2$. Exclusion of {\it u} and {\it z}$^\prime$-band data (red distribution) results in a dearth of identifications at $1<z<2.5$ and excess at $z<0.5$ and $z\sim3$.}
\label{fig:z_dist_and_compare}
\end{figure}

\begin{figure}
\includegraphics[width=\columnwidth]{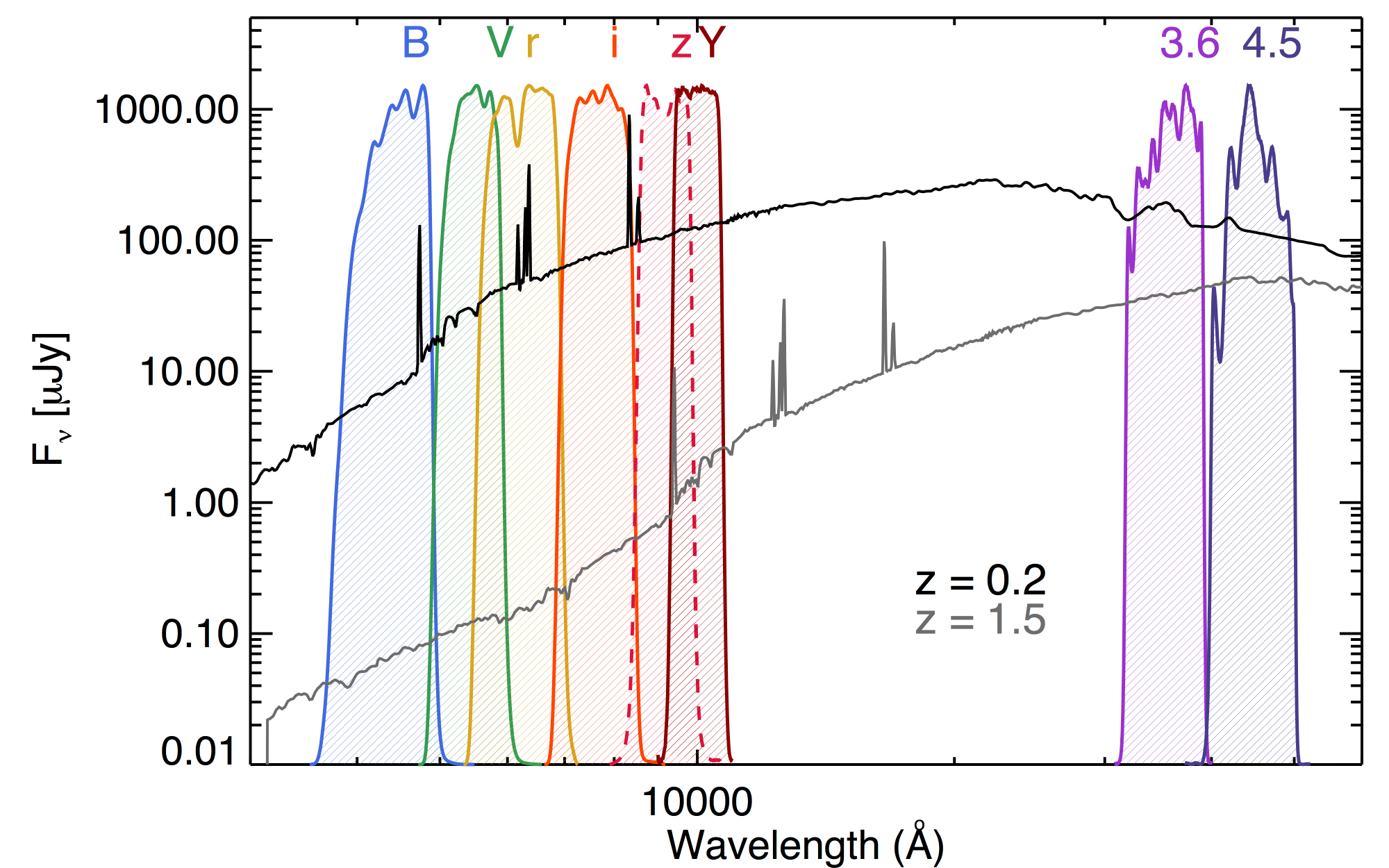}
\vspace*{-5mm}
\caption{Transmission profiles of Subaru optical ($BVrizY$) and {\it Spitzer} near-IR (3.6 and 4.5\,$\micron$) photometric filters. SEDs of sources with $z=0.2$ and $z=1.5$ are shown in black and gray respectively. The $z^\prime$-band profile is dashed to represent the missing $z^\prime$-band coverage in half of the field. Lack of UV and near-IR coverage makes the accurate identification of $z\sim1.5$ sources particularly challenging.}
\label{fig:filter_profiles}
\end{figure}

Figure \ref{fig:z_dist_and_compare} illustrates how missing photometric information affects {\sc EAZY}'s ability to generate accurate photometric redshifts. From this we can see where the estimated photometric redshifts in the SuperCLASS sample might fail. Most noticeably, it struggles to identify $1<z<2$ sources. Prominent features in the modeled spectra, e.g. the Lyman and Balmer breaks at 912\,\AA\ and 3646\,\AA\ as well as the 4000\,\AA\ break, facilitate a more robust identification for {\sc EAZY} to recover the most accurate redshift estimates. Without observations in the \textit{u}-band, we are missing the Lyman break and Lyman-$\alpha$ emission (1216\,\AA) at $z<2$. Similarly, without $z^\prime$-band, we miss the Balmer break from $1.3<z<1.5$ and the 4000\,\AA\ break from $1<z<1.7$. The OIR filter transmission profiles are shown for the bands we have observed in Figure \ref{fig:filter_profiles}. Two SEDs are shown for a $z=0.2$ and $z=1.5$ galaxy illustrating where these distinct emission and absorption features lie.

\begin{figure}
\includegraphics[width=\columnwidth]{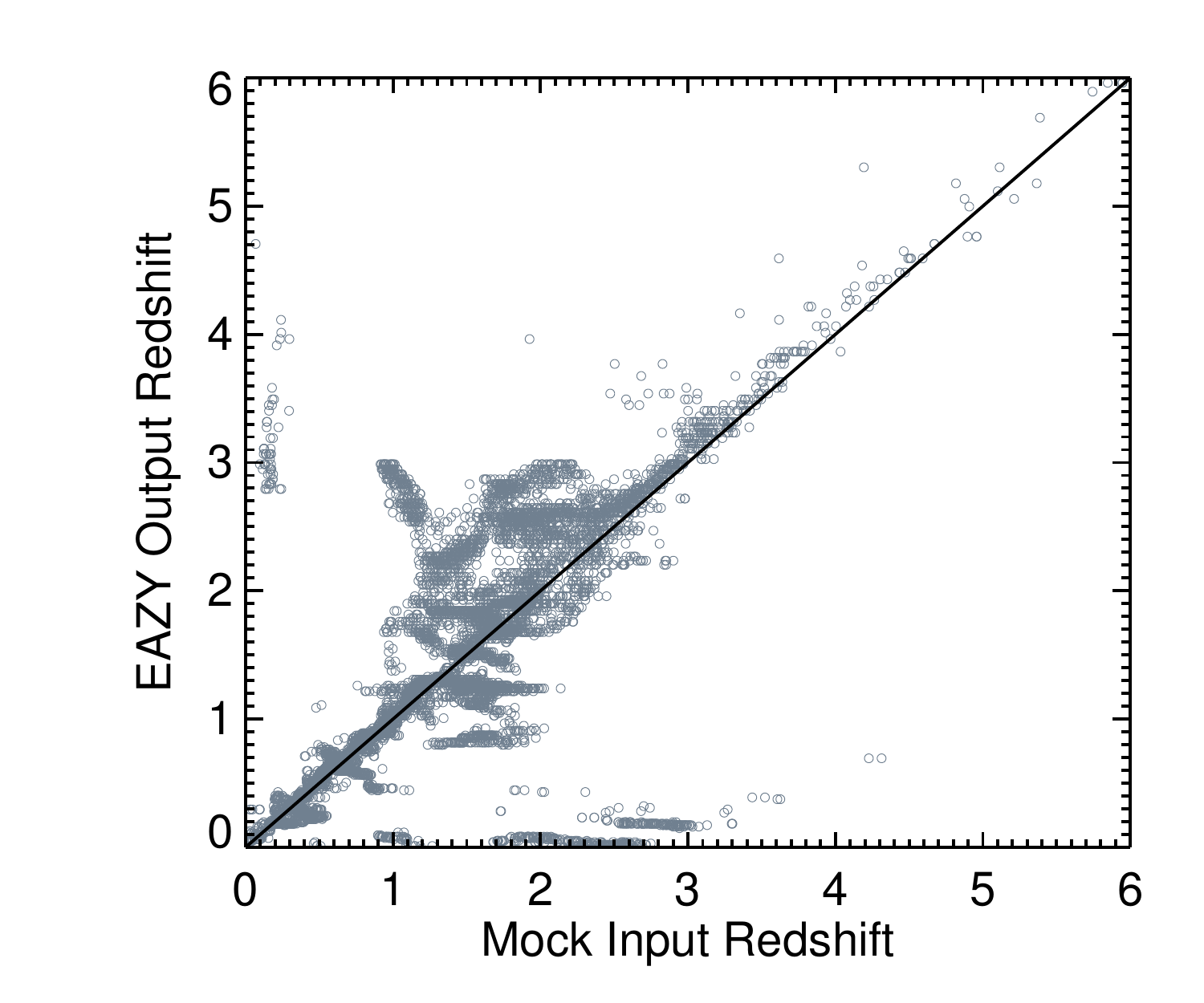}
\vspace*{-5mm}
\caption{EAZY output photometric redshift versus the input redshift from the mock catalog using {\it BVriY}, 3.6\,\micron, and 4.5\,\micron\ photometry. The most common failure modes are $z\sim1.5$ sources misidentified at $z\sim1$ and $z\sim3$.}
\label{fig:eazyz_vs_mockz}
\end{figure}

To get a better understanding of how sources might be misidentified by {\sc EAZY}, we plot the photometric redshift determined by {\sc EAZY} versus the input redshift from the mock catalog in Figure \ref{fig:eazyz_vs_mockz} using the star-forming (BC03) templates. This version of the mock catalog incorporates the {\it BVriY}, 3.6\,\micron, and 4.5\,\micron\ filters, excluding $z^\prime$-band as only half of the DR1 sample has $z^\prime$-band coverage. This combination of filters matches those used in making the red distribution in Figure \ref{fig:z_dist_and_compare}. Here we can see where the biggest disparities occur in our photometric redshift estimates, including many $1<z<2$ sources being misidentified at $2<z<3$. Furthermore, a non-negligible fraction of $1<z<3$ mock galaxies are identified as $z<0.5$ sources. 

We find that close visual inspection of the real data and their galaxy SEDs is the most effective quality check and identify 3\% (149/154) of sources with poor photometric redshift fits as stated in Section \ref{subsec:phot-z_fitting}.

\section{Results}
\label{sec:results}

\subsection{Redshift Distribution}
\begin{figure}
\includegraphics[width=0.5\textwidth]{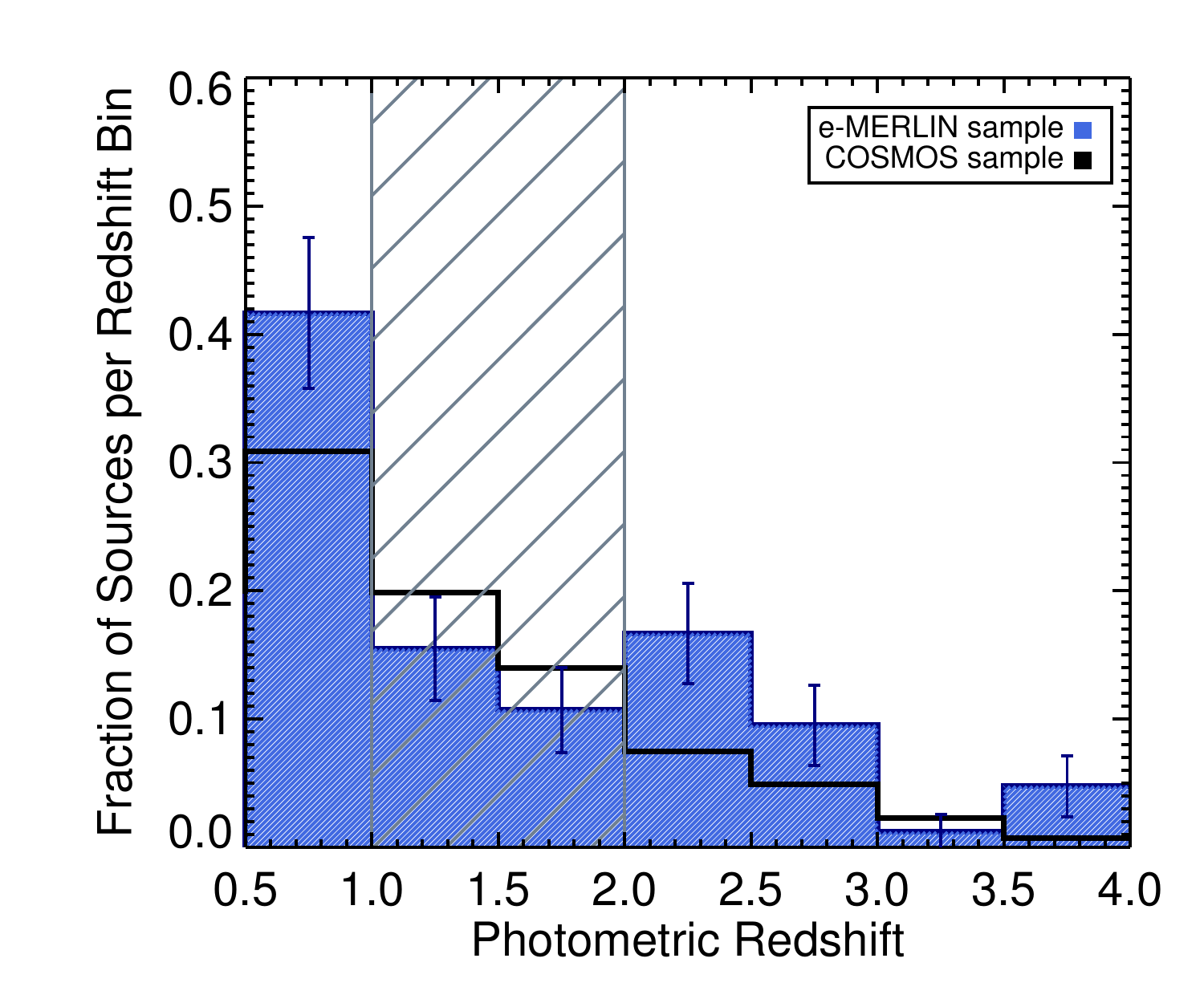}
\vspace*{-5mm}
\caption{Fraction of sources per redshift bin of \eMERLIN\ sources with OIR counterparts ($i^\prime<24.5$) (blue histogram) compared to the redshift distribution of our comparative sources from COSMOS (black). We have omitted the first redshift bin ($0<z_{\rm phot}<0.5$) as it is known to have an excess of sources due to the Abell clusters in the field. The gray hashed region ($1<z_{\rm phot}<2$) shows where we find the largest absence of sources due to our current photometric coverage. The excess at $0.5<z<1$ is likely boosted by sources sitting at $1<z<3$ as suggested by our simulations shown in Figure \ref{fig:eazyz_vs_mockz}.}
\label{fig:zdist_emerlin}
\end{figure}

Here we present the positions of the 149 \eMERLIN\ sources with optical counterparts (matched to OIR counterparts within $<1\arcsec$), their photometric redshifts, $i^{\prime}$-band magnitudes and associated errors, \eMERLIN\ flux densities, and VLA radio luminosities (See Table \ref{tab:result_table}). These galaxies span a range of redshifts ($0.2<z_{\rm phot}<3.6$) and typically have photometric data from seven of eight OIR filters. Figure \ref{fig:zdist_emerlin} shows the photometric redshift distribution and bootstrap errors of sources with $z_{\rm phot}>0.5$. We compare this distribution to that of the COSMOS sources which were used as priors in Section \ref{subsec:cosmos_priors}. Finally, we choose to omit the lowest redshift bin as it is known to have an excess of sources due to the presence of the Abell clusters. We see a slight dearth sources between $1<z<2$ as well as an excess of sources between $0.5<z<1$, $2<z<3$, and $z>3.5$. While our limited photometry may result in some ambiguous redshift identifications (as illustrated in Figure \ref{fig:sed_compare}), this was only the case for $<10\%$ of sources.

\subsection{Radio Luminosities}
\label{sec:results:radio_lum}
\begin{figure}
\includegraphics[trim={0.3in 1.5in 1.2in 1.8in},width=\columnwidth]{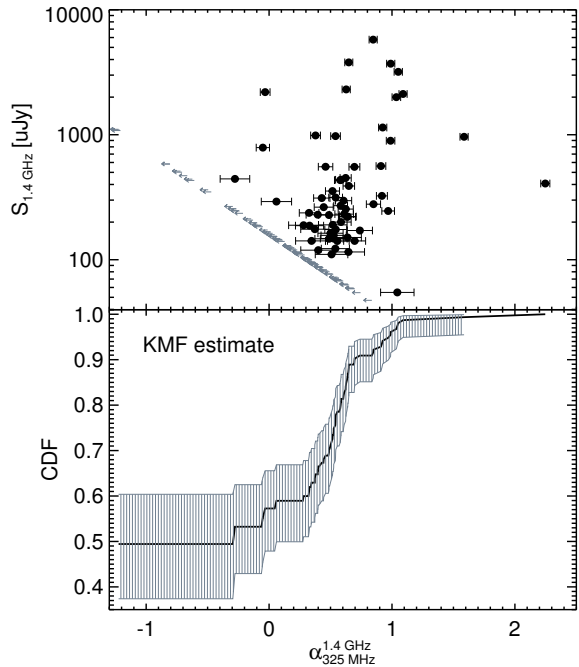}
\vspace*{-5mm}
\caption{{\it Top:} 1.4\,GHz VLA flux density versus spectral index for galaxies with both VLA and GMRT detections. Gray arrows show spectral index upper limits for sources assuming a $5\,\sigma$ GMRT detection (170\,\uJy.) {\it Bottom:} CDF of calculated spectral indices and upper limits, determined by the Kaplan-Meier estimator.}
\label{fig:alpha_dist}
\end{figure}

Of the 149 sources, we determine spectral indices ($\alpha$) for 60 that possess radio detections from both VLA and GMRT (325\,MHz). In this paper we adopt the convention $S\propto \nu^{-\alpha}$ where $\alpha>0$. Figure \ref{fig:alpha_dist} shows the 1.4\,GHz VLA flux densities versus $\alpha$. Errors are propagated given the scatter in flux density. For the remaining galaxies lacking GMRT detections, we determine upper limits on the spectral index based on their VLA detections and the 5\,$\sigma$ detection limit for GMRT (170\,\uJy\ beam$^{-1}$). The spectral index upper limits are denoted as gray arrows. Due to the large number of upper limits, we employ the Kaplan-Meier estimator modified to handle left-censored data \citep{cameron_davidson_pilon_2019_3386382} to produce a cumulative distribution function (CDF) and estimate the median of the combined data and upper limits. Based on this, we can constrain the median of the full sample to be $\langle \alpha \rangle \geq $ --0.28 and 80\% of the sample has $\alpha<0.59$. This median value is significantly shallower than spectral indices of star-forming galaxies in the literature. For example, \citet{2017MNRAS.469.3468C} find $\langle \alpha \rangle=0.74^{+0.27}_{-0.41}$ over the same frequency range. However, our calculated value is driven by the lower limits and we would expect the distribution to shift towards steeper slopes if more GMRT observations became available. The median of the constrained sub-sample (excluding the upper limits) is $\langle \alpha \rangle=0.59\pm0.04$. This is not representative of the full sample, but it is well within the lower limits of the median value from the literature. 

\begin{figure}
\includegraphics[width=\columnwidth]{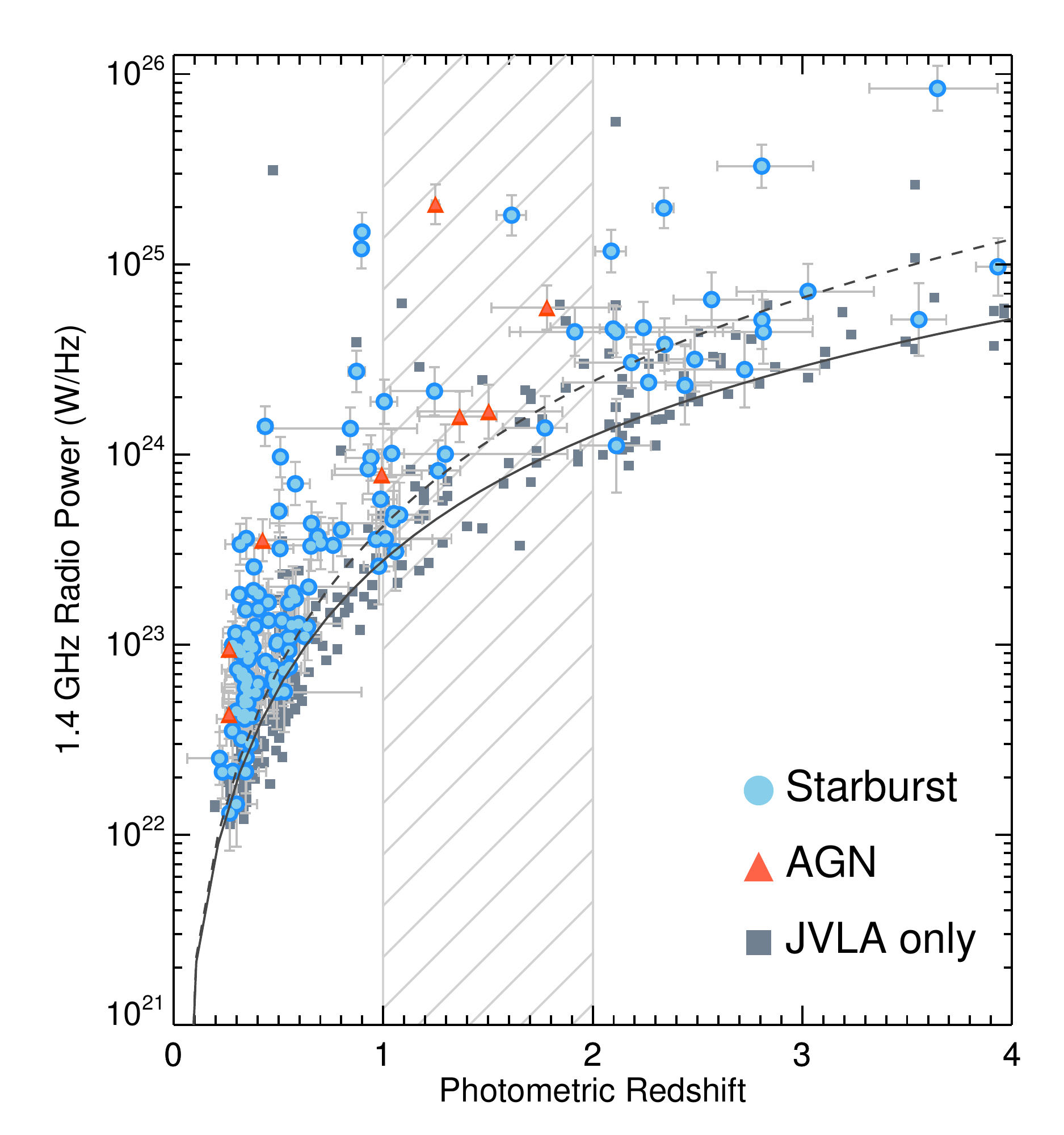}
\vspace*{-5mm}
\caption{Radio power of the 149 SuperCLASS sources with \eMERLIN\ and optical/near-IR detections. The broad range of radio power suggests the presence of both star-forming galaxies and AGN. Blue circles represent sources fit with the BC03 star forming templates while orange triangles are sources fit with the {\it XMM} AGN templates. Gray squares indicate VLA} sources with no \eMERLIN\ counterpart. Dark gray lines show selection curves for 5\,$\sigma$ \eMERLIN\ detections assuming a spectral index $\alpha$ of 0.5 (solid line) and 1.1 (dashed line) corresponding to the rough minimum and maximum spectral index of diffuse synchrotron emission from cosmic ray electrons \citep{Lisenfeld2000}. The gray hatched region replicates the redshift range where we see a modest dearth of objects detected in the full catalog due to the current photometric band coverage ($1<z_{\rm phot}<2$).
\label{fig:radio_lum}
\end{figure}

Figure \ref{fig:radio_lum} shows the rest-frame 1.4\,GHz radio luminosities of all 149 \eMERLIN\ DR1 radio sources. This sample is detected with both \eMERLIN\ and VLA and we chose to adopt the unresolved VLA flux densities as the total flux given the possibility that \eMERLIN\ may resolve out diffuse emission. VLA sources without \eMERLIN\ counterparts are also shown for comparison.

We convert the VLA flux densities and associated errors to rest frame radio luminosities and assume a synchrotron spectral index of $\alpha=0.7$ for sources lacking GMRT detections, as synchrotron radiation is expected to dominate radio emission for most \uJy\ radio sources at frequencies below 30\,GHz \citep{Condon1992}. Sources which best fit star-forming templates are shown in blue while orange denotes sources best fit by AGN templates. The median $z_{\rm phot}$ and L$_{\rm rad}$ of the star-forming versus AGN sources are comparable ($\langle z_{\rm phot}\rangle=0.53$, $\langle$L$_{\rm rad}\rangle=2.43\times10^{23}\,$W\,Hz$^{-1}$ for star-forming galaxies and $\langle z_{\rm phot}\rangle=0.47$, $\langle$L$_{\rm rad}\rangle=2.17\times10^{23}\,$W\,Hz$^{-1}$ for AGN). Our sample spans a broad range of radio power, covering five orders of magnitude. We plot the 149 sources which have photometric redshift uncertainty $\Delta z / (1+z) <1$ and find that 8 out of 149 sources ($\sim$\,5\,$\%$) are better fit by the AGN templates. This is lower than AGN fractions in the literature \citep{Laird2010,Georgantopoulos2011,Johnson2013,Wang2013}, however we have adopted a conservative approach for identifying AGN via photometric redshift fits. We will explore the AGN fraction more fully in a future paper.

If dominated by star formation, we expect these \uJy\ radio sources to correspond to FIR luminosities of LIRGs ($10^{11}L_{\odot}$) and ULIRGs ($10^{12}L_{\odot}$) as we step up in radio power beyond $\sim$\,6$\times10^{22}$ W\,Hz$^{-1}$ at $z\sim0.25$ (LIRGs) and $\sim\,$6$\times10^{23}$ W\,Hz$^{-1}$ at $z\sim0.8$ (ULIRGs) respectively. Our sample follows the same progression and extends to higher redshift than previous detailed studies of \uJy\, radio sources \citep{Chapman2005,Barger2007}.

\begin{table*}
\caption{Photometric Redshifts and Radio Properties of \eMERLIN\ \uJy\ Radio Sources}
\label{tab:result_table}
\begin{tabular}{cccccc}
\hline
\eMERLIN\ Name & $z_{\rm phot}$ & $i^\prime$ Magnitude & S$_{\rm 1.4\,GHz}$ & L$_{\rm rad}$ & Template \\
 & & [AB] & [\uJy] & [W\,Hz$^{-1}$] & \\
\hline 
SCL-EM-J102355.63+675349.58 & 0.63$^{+0.03}_{-0.05}$ & 21.51$\pm$0.01 & 120$\pm$12 & (1.10$\pm0.19)\times10^{23}$ & BC03 \\
SCL-EM-J102155.19+675633.39 & 1.88$^{+0.43}_{-0.11}$ & 22.51$\pm$0.01 & 106$\pm$14 & (2.03$\pm0.16)\times10^{24}$ & BC03 \\
SCL-EM-J102609.28+675935.49 & 0.33$^{+0.04}_{-0.04}$ & 19.23$\pm$0.01 & 110$\pm$13 & (6.75$\pm0.13)\times10^{22}$ & BC03 \\
SCL-EM-J102608.87+675755.47 & 2.34$^{+0.23}_{-0.03}$ & 22.71$\pm$0.01 & 952$\pm$27 & (1.98$\pm0.11)\times10^{25}$ & BC03 \\
SCL-EM-J102545.40+675636.41 & 0.45$^{+0.04}_{-0.03}$ & 20.92$\pm$0.01 & 279$\pm$8 & (1.34$\pm0.14)\times10^{23}$ & BC03 \\
SCL-EM-J102651.62+680334.45 & 1.53$^{+0.08}_{-0.43}$ & 22.09$\pm$0.01 & 228$\pm$12 & (2.76$\pm0.13)\times10^{24}$ & BC03 \\
SCL-EM-J102609.88+680517.22 & 1.78$^{+0.16}_{-0.12}$ & 23.82$\pm$0.01 & 267$\pm$12 & (5.91$\pm0.12)\times10^{24}$ & XMM \\
SCL-EM-J102520.92+680808.33 & 0.87$^{+0.02}_{-0.02}$ & 21.72$\pm$0.01 & 715$\pm$33 & (2.73$\pm0.11)\times10^{24}$ & BC03 \\
SCL-EM-J102150.75+680558.76 & 0.23$^{+0.01}_{-0.01}$ & 19.35$\pm$0.01 & 124$\pm$12 & (2.14$\pm0.14)\times10^{22}$ & BC03 \\
SCL-EM-J102714.51+681032.67 & 3.03$^{+0.15}_{-0.18}$ & 22.21$\pm$0.01 & 201$\pm$10 & (7.19$\pm0.14)\times10^{24}$ & BC03 \\
... & ... & ... & ... & ... & ... \\
\hline
\end{tabular}

\begin{flushleft}
{\bf Note --} Photometric redshift estimates after applying COSMOS priors, $i^\prime$-band magnitudes, \eMERLIN\ 1.4\,GHz flux densities, and VLA 1.5\,GHz radio luminosities of \uJy\ radio sources in the SuperCLASS DR1 field. The template column denotes whether the photometric redshifts were better fit using the star-forming (BC03) or the AGN ({\it XMM}) templates. The full table consisting of all 149 sources and complete photometric information will be available online.
\end{flushleft}
\end{table*}

\subsection{$z\sim0.2$ Cluster Map}
\begin{figure*}
\centering
\includegraphics[scale=0.8]{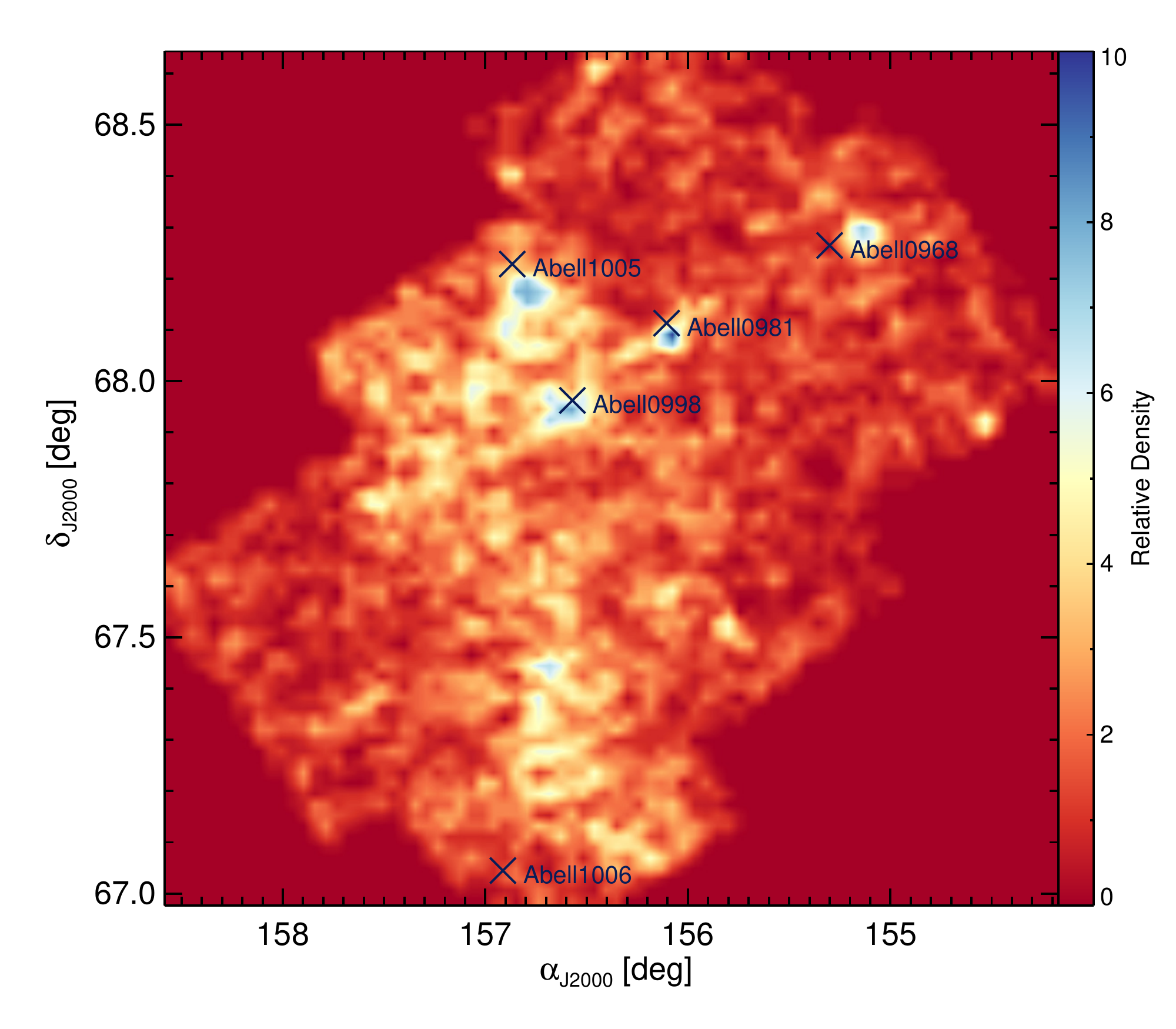}
\vspace*{-7mm}
\caption{Spatial density of galaxies with $z_{\rm phot}=0.2\pm0.08$ and $i^\prime<22.5$. Dark blue crosses are centered on the Abell cluster positions from the literature. We count the number of sources within 75\arcsec\ diameter cells and divide by the mean density across the whole field, resulting in the relative density of sources shown in this figure.}
\label{fig:cluster_map}
\end{figure*}

The five Abell clusters in the SuperCLASS field were observed as part of the {\it ROSAT} All--Sky Survey \citep{Voges1999}. X-ray observations for Abell 968, 981, 998, and 1005 were taken by the Position Sensitive Proportional Counter during the pointed GO phase of the {\it ROSAT} mission \citep{David1999}; the ROSAT positions are marked in Figures \ref{fig:SC_map} and \ref{fig:cluster_map}. An X-ray luminosity measurement for Abell 1006 as well as richness for all five clusters is found in \cite{Briel1993}. Masses can be estimated from the $L_X-M_{500}$ relation as discussed in \cite{Peters2018}.

\subsubsection{Abell Cluster Discussion}
\label{subsub:cluster_discussion}
The Abell catalog of rich galaxy clusters \citep{Abell1958,Abell1989} is an invaluable repository of positions and redshifts of over 4000 galaxy clusters. However, not all of the galaxy clusters listed within \citet{Abell1989} are confirmed. The catalog is based on purely visual surveys of apparent areal densities of galaxies and it is clearly stated the list should not be taken as a definitive list of clusters, but rather a finding list of apparent rich clusters which need further investigation. Furthermore, given the high galactic latitude of the SuperCLASS field, it is unsurprising that information and observations for Abell 968, 981, 998, 1005, and 1006 is lacking.

Redshifts for clusters within the Abell catalog are an average of all spectroscopic redshifts within the determined counting radius. \citet{Struble1999} report Abell 968 and 981 have a recorded redshift for only one source, two sources were averaged for Abell 1005, three for Abell 1006, and five sources were averaged for Abell 998. Thus, the cluster redshifts are not well constrained and while the galaxies within them may be spatially concentrated, it is possible that the previously identified spectroscopic sources are not associated with the cluster that could sit at a different redshift. For this reason, we expand our $z_{\rm phot}$ cut to $0.12<z_{\rm phot}<0.28$ as stated in subsection \ref{subsub:cluster_recovery}, otherwise we simply do not recover a significant number of sources to constitute a cluster based on the Abell method (at least 30 sources).

\subsubsection{Cluster Recovery and Analysis}
\label{subsub:cluster_recovery}
We create a spatial density map of galaxies with photometric redshifts consistent with the reported $z\approx0.2$ Abell cluster redshifts to investigate whether or not we recover the overdensities spatially using our full photometric redshift catalog. To allow for even source density across the field which has variable depth, and to ensure robust photometric redshifts, we make a magnitude cut for sources with an \textit{i}$^\prime$-band magnitude $<22.5$ and $z_{\rm phot}=0.20\pm0.08$. The magnitude cut was made to minimize contamination from higher redshift targets with poorly constrained redshifts. We discuss the decision to use this broader range of redshifts even though cluster redshifts are all reported to be within $z=0.2\pm0.005$ in Section \ref{subsub:cluster_discussion}. We divide the field into a grid consisting of $75\times75\arcsec$ cells and measure the density of sources within each cell. This is then divided by the mean number of sources per cell across the entire field, resulting in the relative overdensity. We explored several different grid spacings, but settled on 75\arcsec\ as it provides a smooth distribution of sources across the map and the density of well-constrained sources is high enough to discern overdense environments.

Figure \ref{fig:cluster_map} shows that we successfully recover Abell 968, 981, 998 and 1005 at 7, 9, 7, and 8$\,\sigma$ significance respectively, though all are slightly offset from the reported cluster centers based on X-ray emission. We note that there is a spatial offset between ROSAT positions and the peak galaxy densities in our density map. While this could possibly be physical -- whereby interactions in the super-cluster environment lead to shock-heated ICM gas offset from the dark matter potentials -- a more likely explanation in this context is imprecise astrometric constraints from ROSAT given the 0.5\arcmin\ beam and few photons used for identification. We also identify one additional high-density region with 6\,$\sigma$ significance at ($\alpha,\delta$)=(156.72,67.45) that is not detected in the existing X-ray data from {\it ROSAT} (0.1--2.4\,keV) or {\it Suzaku} (0.2--600\,keV). 

As determined by our mock catalog analysis (Section \ref{sec:mocks}), there is the potential for an excess of sources to lie at $z_{\rm phot}<0.5$ due to the current combination of available photometric filters. We create density maps across several redshift slices from $0<z_{/rm phot}<0.5$ to determine if the apparent spatial clustering around the cluster positions may be caused by poor photometric redshift fits and projected interlopers rather than the actual presence of a cluster. No redshift slice other than those encompassing our selected range of $0.12<z_{\rm phot}<0.28$ produces an overdensity around the distinct cluster positions. The relatively even distribution of sources across the field suggests that even with the inclusion of some poorly fit interlopers, they would not effectively inflate or alter our cluster detections.

\begin{figure}
\includegraphics[width=\columnwidth]{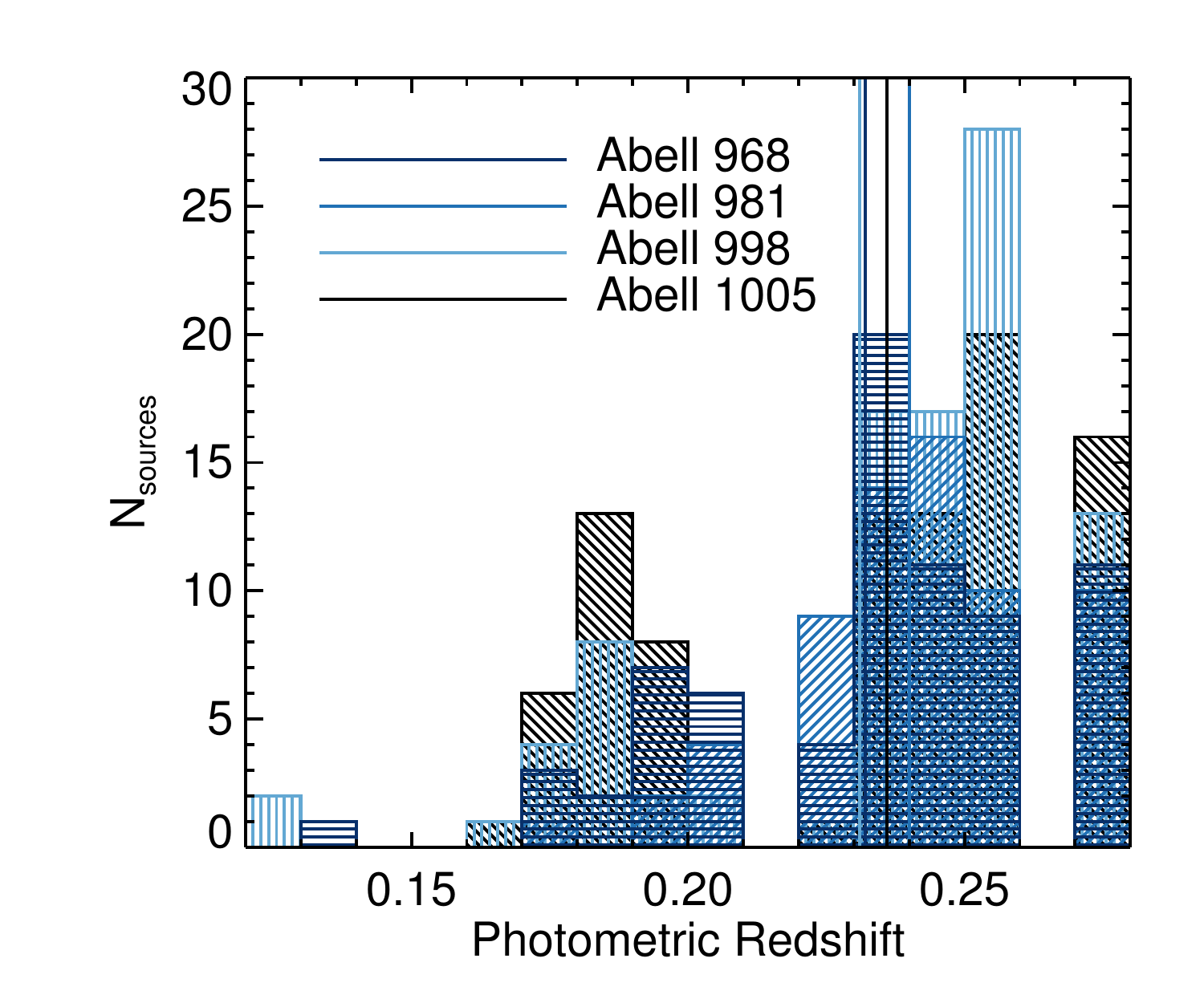}
\vspace*{-5mm}
\caption{Photometric redshift distribution of sources within 3\arcmin\ of the cluster centroids determined from Figure \ref{fig:cluster_map}. The average $z_{\rm phot}$ of sources around each cluster is listed in Table \ref{tab:abell} and shown as vertical lines here.}
\label{fig:cluster_z_dist}
\end{figure}

We calculate the average cluster photometric redshifts based on sources found within 3\arcmin\ of the cluster centroids determined from the density map (Figure \ref{fig:cluster_map}). The redshift distribution for each cluster is shown in Figure \ref{fig:cluster_z_dist}. Errors are quoted as the standard deviation of the peak in the distribution for each cluster. Table \ref{tab:abell} summarizes characteristics of the known clusters from the literature and our findings. 

\begin{figure}
\includegraphics[width=\columnwidth]{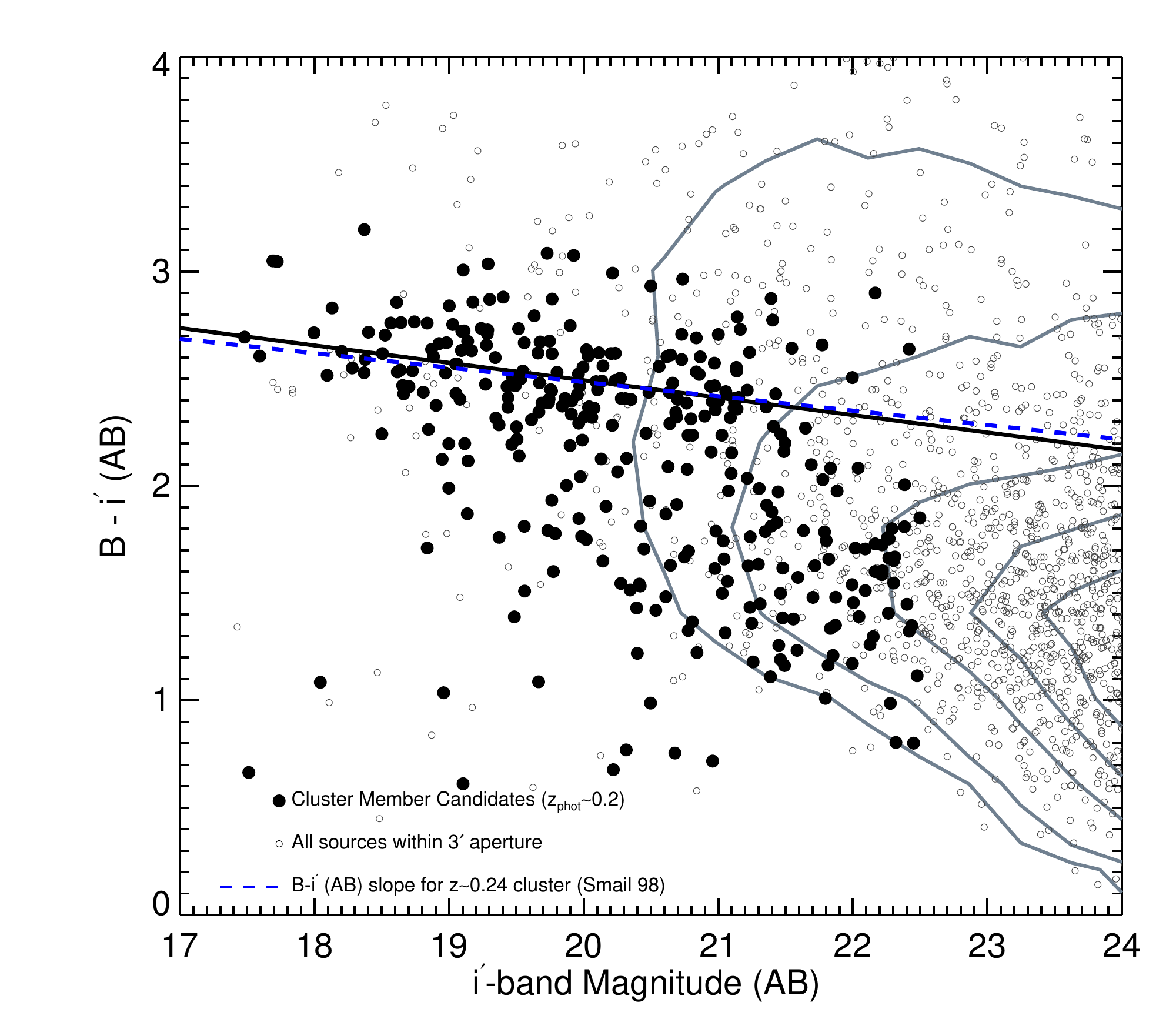}
\vspace*{-5mm}
\caption{$B-i^{\prime}$ versus $i^{\prime}$ color-magnitude plot of cluster member candidates and SuperCLASS field sources. Cluster member candidates consist of sources found within 3\arcmin\ of the new measured cluster positions based on the density map with $0.12<z_{\rm phot}<0.28$ (filled circles) and open circles are the remaining sources withing the aperture. Gray contours show the color-magnitude distribution for the entire field. For comparison we plot the red sequence slope of a $z=0.24$ galaxy cluster (blue dashed line) determined in \citet{Smail1998} which has been converted from Vega to AB magnitudes.}
\label{fig:color-mag}
\end{figure}

A color-magnitude diagram is shown in Figure \ref{fig:color-mag}. We define cluster member candidates as sources within 3\arcmin\ of the measured cluster centroids derived from the density map based on the photometric redshifts between $z_{\rm phot}=0.2\pm0.08$. We center the 3\arcmin\ apertures on the cluster positions of Abell 968, 9181, 998, and 1005 as they are the largest apparent structures. It is still highly likely that we are including non-cluster members along the line of sight using this method. Sources with $z_{\rm phot}$ within the stated range are shown as black filled circles and referred to as cluster member candidates. Open circles show remaining sources found within the 3\arcmin\ aperture, but with photometric redshifts inconsistent with the cluster redshifts. We also compare the $B-i^\prime$ versus $i^\prime$ colors of the cluster candidates to all the remaining sources in the SuperCLASS field (gray contours). As expected, galaxies thought to be cluster members lie within a narrower range of colors, along a red sequence of passive galaxies, compared to the broad range of the full sample. We fit a line to sources that appear to fall along this red sequence ($18<i^{\prime}$-band$<22.5$ and $B-i^\prime>2$) and compare it to the slope determined in \citet{Smail1998} for a $z=0.24$ galaxy cluster (converted from Vega to AB for a direct comparison). Determining the existence of these five clusters still requires follow-up spectroscopic observations to determine accurate redshifts. 

\begin{table*}
\footnotesize
\caption{Observational Properties of Abell Clusters in SuperCLASS}
\label{tab:abell}
\begin{tabular}{cccccccccc}
\hline
Cluster & RA$_{\rm J2000}$ & DEC$_{\rm J2000}$ & RA & DEC & $z_{\rm spec}$ & $z_{\rm phot}$ & M$_{500}$ & L$_{0.1-2.4\rm keV}$ & Rel. Density \\
Name & (literature) & (literature)& (this work) & (this work) & & (this work) & [M$_\odot$] & [erg s$^{-1}$] & N$_{\rm cell}$/$\langle$ N$_{\rm cells} \rangle$ \\
\hline
Abell 968* & 10:21:09.5 & +68:15:53 & 10:20:31.20 & +68:17:24.00 & 0.195 & 0.232$\pm$0.029 & (1.2$\pm$0.3)$\times10^{14}$ & 0.401$\times10^{44}$ & 7 \\
Abell 981* & 10:24:24.8 & +68:06:47 & 10:24:19.20 & +68:05:23.98 & 0.201 & 0.240$\pm$0.021 & (2.7$\pm$0.7)$\times10^{14}$ & 1.670$\times10^{44}$ &  9 \\
Abell 998* & 10:26:17.0 & +67:57:44 & 10:26:19.20 & +67:56:59.99 & 0.203 & 0.231$\pm$0.034 & (1.2$\pm$0.3)$\times10^{14}$ & 0.411$\times10^{44}$ &  7 \\
Abell 1005* & 10:27:29.1 & +68:13:42 & 10:27:07.19 & +68:10:48.00 & 0.200 & 0.236$\pm$0.033 &  (1.0$\pm$0.2)$\times10^{14}$ & 0.268$\times10^{44}$ &  8 \\
Abell 1006 & 10:27:37.2 & +67:02:41 & - & - & 0.204 & - &  (2.4$\pm$0.6)$\times10^{14}$ & 1.320$\times10^{44}$ &  0 \\
\hline
\end{tabular}

\begin{flushleft}
{\bf Note --} Cluster information for the five Abell clusters located within the SuperCLASS field. Positions and X-ray luminosities are taken from the BAX database \citep{Sadat2004}. Redshifts and mass estimates have been collected from \cite{Peters2018}. We include the measured centroids from the photometric redshifts as well as the average photometric redshift for each cluster. Errors are quoted as the standard deviation of the N($z$) distribution for each cluster. 
The relative density is the number of sources within a $75\times75\arcsec$ cell divided by the mean number of sources per cell obeying the following constraints, \textit{i}$^\prime$-band magnitude $<22.5$ and $z_{\rm phot}=0.20\pm0.08$, across the entire field. *Abell 968, 981, 998, and 1005 are recovered in our spatial density map (Figure \ref{fig:cluster_map}).
\end{flushleft}
\end{table*}

\section{Summary}
\label{sec:summary}
We present photometric redshifts determined from Subaru ({\it BVrizY}) and {\it Spitzer} (3.6, 4.5\,$\micron$) photometry of 370,000+ sources (176,000+ with $i^\prime< 24.5$) across the SuperCLASS $\sim$\,2\,deg$^2$ field. The SuperCLASS survey will act as a testbed for weak lensing studies as well as a blind search for DSFGs to study their morphological properties from optical through radio wavelengths. We use the 1.4\,GHz radio sources in the COSMOS catalog with similar VLA flux densities and 3.6\,\micron\ magnitudes, and well-constrained photometric redshifts as priors to weight the PDFs generated by EAZY and further constrain our redshift estimates. Having photometric redshifts is crucial for this survey analysis. Beyond the radio weak lensing goals, we can now more clearly characterize the \uJy\ radio source sample, examine how they may form and evolve across cosmic time, in addition to constraining the source density of these systems.

We also present \eMERLIN\ 1.4\,GHz radio luminosities for the first 149 \eMERLIN\ radio sources with optical and near-IR counterparts and robust photometric redshifts in the SuperCLASS DR1 (0.26\,deg$^2$) field. We examine the \uJy\ radio source population in the DR1 survey area, their radio luminosities, and have determined whether the sources are better fit by star-forming or AGN SED templates to account for the expected variation in the types of radio sources in the field. Of the 149 sources, we find 141 starburst galaxies and 8 AGN, all falling within a redshift range of $0<z<4$ and radio luminosities of $10^{21}-10^{25}$\,W\,Hz$^{-1}$. We caution that there are likely more AGN in the sample based on literature analysis of the \uJy\ radio galaxy population, but leave further analysis of the AGN content -- based on photometry and a morphological analysis -- to a future work.

Based on the mock observations we expect some uncertainty in our photometric redshifts and see a slight dearth of sources at $1<z<2$ in the full catalog due to our current wavelength coverage, but plan to rectify this with future observations. Increasing the photometric coverage of the field will reduce the potential to miss easily identifiable spectral features. Obtaining observations in \textit{u}-band ($\sim$3000\,\AA--4000\,\AA) would allow us to observe the Lyman break from $2.2<z<3.2$, providing better identification of higher-$z$ objects as well as Lyman-$\alpha$ emission at $z<2$. We also believe $z^\prime$-band would be helpful in identifying the Balmer break from $1.1<z<1.5$. Future improvements to the SuperCLASS catalog may come with the addition of observations in these photometric bands as well as spectroscopic follow-up of individual sources which can be used as calibrators for the photometric redshifts.

Creating a spatial density map based on photometric redshifts of $0.12<z_{\rm phot}<0.28$ results in the recovery of four (Abell 968, 981, 998, and 1005) out of five suggested Abell clusters in the SuperCLASS field at $>7\,\sigma$. The cluster candidate galaxies occupy a fairly tight and red region of color space, indicating their potential cluster membership and evolution to passive galaxies. However, spectroscopic follow-up is required to confirm the presence of these clusters.

\section*{Acknowledgements}
The authors thank the reviewer, Kenneth Duncan, for his helpful comments and suggestions. SMM thanks the National Science Foundation for support through the Graduate Research Fellowship under Grant No. DGE-1610403. PIs of observations: Casey, Battye, Chapman, Sanders, and Scaife. CMC thanks the National Science Foundation for support through grants AST-1714528 and AST-1814034. SMM and CMC thank the University of Texas at Austin College of Natural Sciences for support. In addition, CMC acknowledges support from the Research Corporation for Science Advancement from a 2019 Cottrell Scholar Award sponsored by IF/THEN, an initiative of Lydia Hill Philanthropies. IH and MLB acknowledge the support of an ERC Starting Grant (grant no. 280127). IH acknowledges support from from the European Research Council in the form of a Consolidator Grant with number 681431, and from the Beecroft Trust. CAH acknowledges financial support from the European Union's Horizon 2020 research and innovation programme under the Marie Sk{\l}odowska-Curie grant agreement No 705332. Based [in part] on data collected at Subaru Telescope, which is operated by the National Astronomical Observatory of Japan. The authors wish to recognize and acknowledge the very significant cultural role and reverence that the summit of Maunakea has always had within the indigenous Hawaiian community. We are most fortunate to have the opportunity to conduct observations from this mountain. This work is based [in part] on observations made with the {\it Spitzer Space Telescope}, which is operated by the Jet Propulsion Laboratory, California Institute of Technology under a contract with NASA. The National Radio Astronomy Observatory is a facility of the National Science Foundation operated under cooperative agreement by Associated Universities, Inc. GMRT is run by the National Centre for Radio Astrophysics of the Tata Institute of Fundamental Research. \eMERLIN\ is a National Facility operated by the University of Manchester at Jodrell Bank Observatory on behalf of STFC. This research made use of data provided by Astrometry.net.





\bibliographystyle{mnras}
\bibliography{smm_papers.bib} 




\bsp	
\label{lastpage}
\end{document}